\documentclass[useAMS]{mn2e}
\usepackage{psfig,epsfig,natbib_sro}

\title[Accretion dynamics in ns-bh binaries]
{Accretion dynamics in neutron star black hole binaries}
\author[S. Rosswog et al.]
       {S. Rosswog$^{1,2}$,  R. Speith$^{3,2}$, G.A. Wynn$^{2}$\\
         $^{1}$ School of Engineering and Science, International University
	 Bremen, Germany\\
	 $^{2}$ Department of Physics and Astronomy, University of Leicester, UK\\
	 $^{3}$ Institut f\"ur Astronomie und Astrophysik, Universit\"at
	 T\"ubingen, Germany }

\date{Accepted YEAR MONTH DAY.
      Received YEAR MONTH DAY;
      in original form YEAR MONTH DAY}

\pagerange{\pageref{firstpage}--\pageref{lastpage}}
\pubyear{2001}

\def\msun{M$_{\odot}$}
\def\be{\begin{equation}}
\def\ee{\end{equation}}
\def\bi{\begin{itemize}}
\def\ei{\end{itemize}}
\def\bea{\begin{eqnarray}}
\def\eea{\end{eqnarray}}
\def\gcc{gcm$^{-3}$}

\begin{document}
\newcommand{\newblock}{}
\maketitle

\begin{abstract}
We perform three-dimensional, Newtonian hydrodynamic simulations with a nuclear equation of
state to investigate the accretion dynamics in neutron star black hole
systems. We find as a general result that non-spinning donor stars yield
larger circularization radii than corotating donors. Therefore, the matter
from a neutron star without spin will more likely settle into an accretion
disk outside the Schwarzschild radius. With the used stiff equation of state 
we find it hard to form an accretion disk that is promising to launch a
gamma-ray burst. In all relevant cases the core of the neutron star survives 
and keeps orbiting the black hole as a mini neutron star for the rest of the
simulation time (up to several hundred dynamical neutron star times scales). 
The existence of this mini neutron star leaves a clear imprint on the
gravitational wave signal which thus can be used to probe the physics at 
supra-nuclear densities. 
\end{abstract}

\begin{keywords}
Accretion; dense matter; stars: neutron; black hole; hydrodynamics; 
methods: numerical 
\end{keywords}

\section{Introduction}
%Why is it interesting?
Neutron star black hole binaries are thought to be excellent candidates for
ground-based gravitational wave (GW) detection (e.g. Belczynski et al. 2002 and
references therein). If their true coalescence rate is 
close to the upper end of the estimated values,  neutron star black hole 
systems may be among the first sources detected. They are further
-together with double neutron star systems- the favoured model for the
subclass of short Gamma-ray bursts (GRBs; Paczynski 1992, Narayan et
al. 1992); a coincident detection of a
gravitational wave ``chirp'' signal together with a GRB could be the ultimate
proof for the compact binary nature of the short GRB progenitors.
Moreover, the cold decompression of nuclear matter by the tidal disruption
of a neutron star has been considered as a possible production
site of rapid neutron capture elements (Lattimer and Schramm, 1974, 1976;
Lattimer et al. 1977 ; Symbalisty and Schramm 1982, Eichler et al. 1989,
Rosswog et al. 1999, Freiburghaus et al. 1999).\\
%What has been done so far?
Full general relativistic simulations of this event are very challenging as
apart from the complicated neutron star physics also the dynamic evolution 
of the space-time has to be followed since the systems of interest have 
comparable masses for both binary components and therefore a fixed space-time
cannot be assumed. But the dynamics of the event is demanding enough
even if treated in the Newtonian approximation. Simulations of neutron star 
black hole encounters have been described in a series of papers by Lee/Lee
and Kluzniak (Lee and Kluzniak 1999a, 1999b, Lee 2000, Lee 2001). They used 
3D smoothed particle hydrodynamics, Newtonian self-gravity plus backreaction 
forces from gravitational wave emission and polytropic equations of state 
(EOS) of varying stiffness. Janka et al. (1999) performed neutron star
black hole merger calculations using a 3D PPM-code, again using Newtonian 
self-gravity  plus backreaction forces, but with the nuclear equation of state 
of Lattimer and Swesty (1991) and a detailed neutrino treatment (Ruffert et
al. 1996).\\
%What is our focus here?
The focus of our investigation in this paper is the complex accretion dynamics 
in a neutron star black hole system. Many of the (analytical) approximations
that are successfully used in other accreting systems are of only limited 
value in this context. For example, the Roche-potential of point mass binaries
is only a rough approximation to the neutron star with its rather flat
density profile; neutron star black hole binaries are most likely not
corotating like other close binaries (Bildsten and Cutler 1992) and, due to
their tiny pressure scale heights in the neutron star surface, 'standard'
estimates of the distance where mass transfer sets in can only give a very
rough guideline.\\
We present in Sect. 2 analytic estimates that explain why
irrotational donor stars do generally form accretion disks easier than
corotating donors. We explain in Sect. 3 our numerical methods and initial
conditions, results concerning the accretion dynamics and the corresponding
gravitational wave emission are presented in Sect. 4. In Sect. 5 we
summarize our results.

\section{Analytic estimates}
%text old version

\begin{figure}
    \hspace*{0cm}\psfig{file=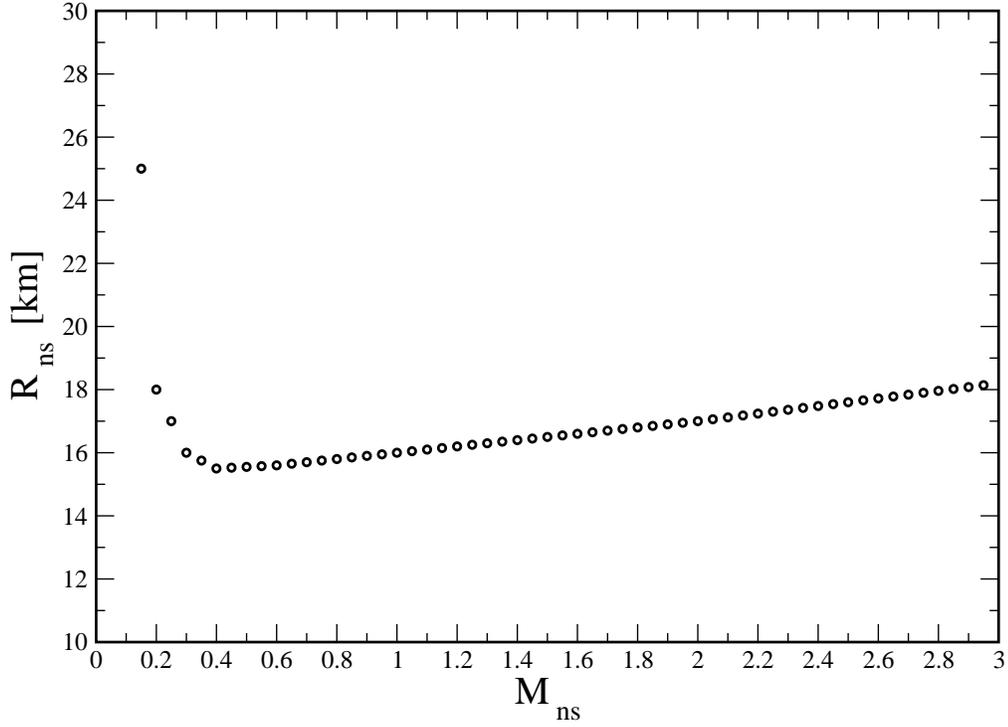,width=12cm,angle=-90}
    \caption{\label{M_R_shen}Mass-radius relationship for the used Shen-EOS:
for neutron star masses above 0.4 \msun the neutron star will shrink upon mass
loss, below this value it increase its radius.}
\end{figure}

While the gravitational wave signature is predominantly determined by the 
bulk motion
of the event, the neutrino and electromagnetic signals from a neutron
star black hole system require an accretion disk to form. Moreover an
accretion disk is thought to be an indispensable ingredient for a GRB central
engine (e.g. Piran 1999, Meszarosz 2002). Therefore we explore with simple
analytical formulae the conditions under which a disk can form.\\
We want to point out the fact that the following estimates only have 
illustrative character since they are very approximate for two reasons:
first, the used formulae refer to Roche-geometry, i.e. to the gravity of 
point masses, while the stiff neutron star equation of state leads to an almost
uniform mass distribution inside the neutron star. Second and more important,
the stiff neutron star equation of state results in tiny pressure scale heights
in the neutron star surface and therefore the level of uncertainty in the 
estimate of the Roche radius can easily correspond to several orders of 
magnitude in terms of pressure scale heights.\\
The dynamics of how the mass transfer proceeds depends crucially on
the way the neutron star reacts on mass loss, i.e. whether it shrinks or 
expands. The mass-radius relationship of a polytropic neutron star with
adiabatic exponent $\Gamma$ is given by (e.g. Kippenhahn 1990)
\be
R \propto M^{(\Gamma-2)/(3\Gamma-4)},
\ee
i.e. for $\Gamma=$ 2 the radius is independent of the mass and for larger 
(smaller) values the neutron star shrinks (expands) on mass loss. For a 
realistic nuclear equation of state $\Gamma$ is density dependent 
(for an example see the comparison between the Shen- and the 
Lattimer-Swesty-EOS in Rosswog and Davies 2002, Figure 5) and therefore the reaction
of the neutron star on mass loss depends on its current mass. The mass-radius
relationship of the used Shen-EOS is given in Fig. \ref{M_R_shen}.
Above 0.4 \msun\ a neutron star will shrink upon mass loss, below this value its radius
increases.\\

\subsection{Accretion disk formation: dependence on donor spin}

\begin{figure}
     \hspace*{0cm}\psfig{file=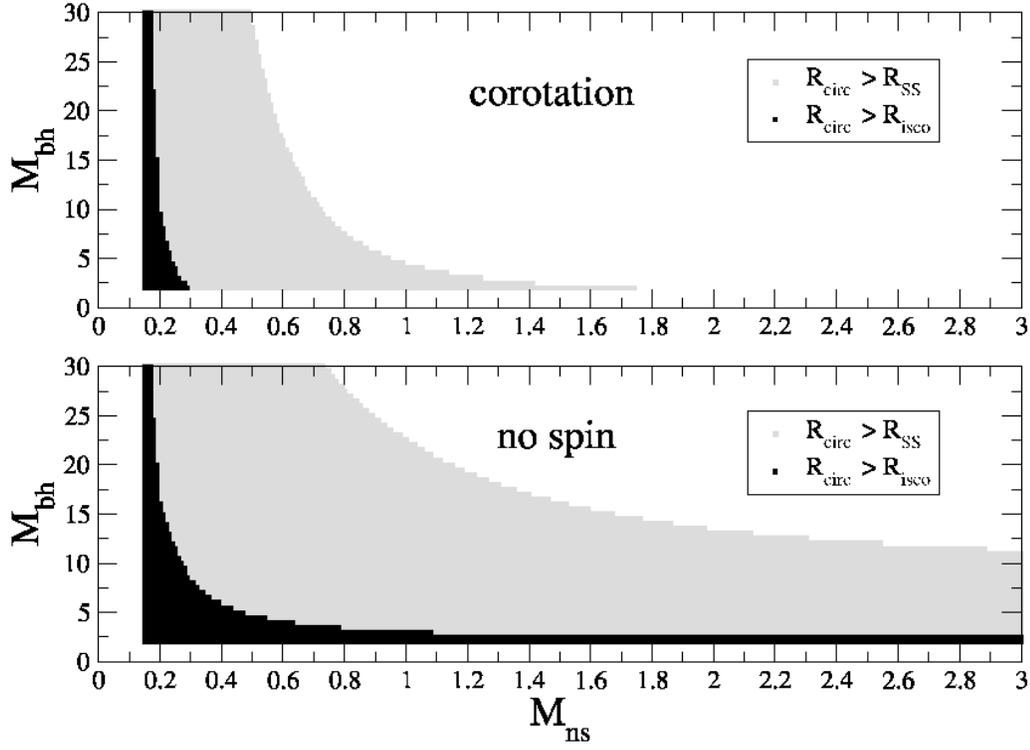,width=12cm,angle=-90}
    \caption{\label{q_accr_disk}Shown are the conditions under which an
            accretion flow from the neutron star will settle into an 
            accretion disk, i.e. $R_{circ} > R_{SS}$ (gray-shaded). We also show 
            (in black) the masses for which the circularisation radius
            is outside the last stable orbit, $R_{isco}=$ 6 M$_{bh}$.
	    The upper panel refers to the case of a corotating donor star,
            while the lower is for a neutron star without spin.}
\end{figure}

We show that -and explain why- it is in general easier to form an
accretion disk due to Roche lobe overflow, if the donor star is non-rotating
rather than being tidally locked. In the non-rotating case, matter flowing
accross the Lagrange point $L_1$ has a larger specific angular momentum and
therefore yields a larger circularization radius than in the corotating case.
As a consequence, in the first case more matter can form an accretion disk
outside the Schwarzschild radius, while for a corotating donor the bulk of
material is fed directly into the hole.\\ 
For the sake of simplicity we assume in the following that the Roche potential
is a valid approximation for our purposes. We further neglect radial
velocities of the binary components and tidal deformations and lag angles 
of the donor star.
Generally, the circularization radius is an estimate for the initial disk
size. It is proportional to the square of the specific angular momentum,
$l_{L_1}$, of the accretion stream at the $L_1$ point as measured in a 
coordinate system ($\Sigma$) with  non-rotating axes and the origin
centred on the black hole, 
\be 
R_{\mathrm circ}= l_{L_1}^2/(G M_{bh}).  \label{Rcirc}
\ee
Therefore, we have to show that the irrotational case yields a larger value
for $l_{L_1}$, i.e.  $l_{L_1}^{irr} > l_{L_1}^{cor}$. The specific angular
momentum is defined as $\vec{l}_{L_1}= \vec{b}_1 \times \vec{v}_{L_1}$, where
$\vec{b}_1$ is the vector from the black hole to the $L_1$ point. 
The velocities at the $L_1$ point are easiest found in a coordinate system 
($\Sigma'$) corotating with the binary centred around the bh.
Here and in the following, all unprimed quantities refer to $\Sigma$, while 
primed ones refer to $\Sigma'$. The coordinate transformation between the two
systems is given by:
\bea 
\vec{x}&=& R_{\varphi}  \vec{x}' \label{trafo_x} \\
\vec{v}&=& \vec{\omega} \times \vec{x} + R_{\varphi}  \vec{v}',\label{trafo_v}
\eea
where $R_{\varphi}$ is the matrix describing a rotation around $\vec{\omega}$. 
The rotation angle $\varphi$ is given by $\omega t$, where $\omega=
|\vec{\omega}|$ is the binary orbital angular velocity.\\
For a corotating donor star all the fluid velocities vanish in the corotating 
frame $\Sigma'$, i.e. $\vec{v}'= 0$. Hence, the velocity at $L_1$ in $\Sigma$ reads
\be 
\vec{v}_{L_1}= \vec{\omega} \times \vec{b}_1
\ee
and therefore
\be 
\vec{l}_{L_1}^{cor}= b_1^2 \vec{\omega}.
\ee
An irrotational donor star, however, appears in $\Sigma'$  to be 
spinning against the orbital motion with $\omega_{spin}= - \omega$,
\be 
\vec{v}_{L_1}'= - \vec{\omega}  \times (\vec{b_1}' - \vec{a}'),\label{v1_irr} 
\ee
where $\vec{a}'$ denotes the vector pointing from the bh to the ns. Inserting
(\ref{v1_irr}) into (\ref{trafo_v}) and considering (\ref{trafo_x}) leads to 
\be 
\vec{v}_{L_1}= \vec{\omega} \times \vec{a}
\ee
and finally
\be 
\vec{l}_{L_1}^{irr}= b_1 a  \vec{\omega}.
\ee
Obviously, as $l_{L_1}^{irr} > l_{L_1}^{cor}$, we have  
\be 
R_{\mathrm circ}^{irr} > R_{\mathrm circ}^{cor}.
\ee
Using (\ref{Rcirc}) and Kepler's third law the circularization radii can be written as
\bea 
R_{\mathrm circ}^{cor}&=& \frac{b_1^4}{a^3} (1+q)\\
R_{\mathrm circ}^{irr}&=& \frac{b_1^2}{a} (1+q),
\eea
where $q= M_{ns}/M_{bh}$.
We now assume that mass transfer sets in once the (unperturbed) neutron star
radius, $R_{ns}$, is comparable to its Roche radius.
The accretion stream from the neutron star to the black hole will result 
in a disk if the stream tries to settle outside the Scharzschild radius of the hole, 
$R_{SS}$, i.e. if $R_{circ} > R_{SS}$. We plot in Fig. \ref{q_accr_disk} the 
conditions under which a disk can form, both for the corotating and the
irrotational case. To produce the plot we have used the formula of Plavec and
Kratochvil (1964) to estimate $b_1$, the Roche radius is approximated
according to Eggleton (1983), and for the radius of the neutron star as a
function of mass we use the (cold, $T=0$) equation of state of Shen et al. 
(1998a,1998b), see Figure \ref{M_R_shen}. The gray-shaded area refers to the
condition $R_{circ} > R_{SS}$ and the black area to $R_{circ} > R_{isco}= 6
M_{bh}$. $R_{isco}$ is the radius of the innermost, stable circular orbit of
a test-particle around a Schwarzschild black hole.\\
Although these are just rough estimates, Fig. \ref{q_accr_disk} 
clearly shows how difficult it is to form an accretion disk 
for a corotating 1.4 M$_{\odot}$ neutron star: only for extremely low-mass
black hole cases (2 M$_{\odot}$) will the accretion stream settle in
a disk around the hole. In the more realistic case of a more massive hole
the accretion stream will be fed directly into the hole and only form 
a disk once the neutron star has expanded on mass loss. In the irrotational
case, however, the fraction of the parameter space that allows for an
accretion disk formation is much larger. From a 1.4 \msun\ neutron star
an accretion disk will form even for a black hole as massive as $\sim
18$ \msun\.\\
All these results are generally confirmed by the set of numerical simulations
presented in Sect. 4.

\section{Model}

\subsection{Numerical Method}
We solve the equations of fluid dynamics in 3D using a Lagrangian particle
scheme (smoothed particle hydrodynamics, SPH; see e.g. Benz 1990, Monaghan 1992). 
For details of the equations (symmetrisation etc.) we refer to the review by
Benz (1990). Rather than the ``standard'' artificial viscosity (Monaghan and
Gingold 1983) we use (Rosswog et al. 2000) time dependent viscosity parameters 
that essentially vanish in the absence of shocks, together with the so-called 
``Balsara-switch'' (Balsara 1995) to suppress spurious viscous forces in pure 
shear flows. For further details of the implementation we refer to Rosswog and
Davies (2002).
The Schwarzschild radius is treated as an absorbing boundary, i.e. particles
that cross the boundary are removed from the simulation. Their mass and linear
momentum are added to the hole in a way that the total centre of mass is conserved.
The microphysics of the neutron star fluid is described using a temperature
and composition dependent nuclear equation of state, based on relativistic mean
field theory (Shen et al. 1998a,b; Rosswog and Davies 2002). Local neutrino
cooling and compositional changes of the neutron star matter due to weak
interactions are accounted for by a detailed, multi-flavour neutrino scheme 
(Rosswog and Liebend\"orfer 2003). Details concerning the neutrino emission
will be presented elsewhere.\\
We use Newtonian self-gravity for the fluid which we evaluate via a binary
tree. The black hole is treated as a Newtonian point mass, results for
pseudo-potentials mimicking relativistic effects will be presented in a
forthcoming paper. For the backreaction onto the fluid flow resulting from the
emission of gravitational waves we use the formalism of Zhuge et al. (1996),
applying the forces only to SPH-particles that have a mass density in
excess of a threshold ($10^{14}$ g cm$^{-3}$), which is above the
phase transition density to neutron star matter. This procedure is
well-justified as in all but one of our simulations a neutron star core 
survives which is, to a good approximation, spherically symmetric. 
In the case of a complete disruption no backreaction force is applied by 
construction. This is also well-justified as the matter distribution for 
this case is very close to axisymmetry.

\subsection{Initial Conditions}
% old text
It is a vital ingredient for these calculations to start out from initial
conditions as accurate as possible since the whole dynamical evolution is
sensitive to the starting configuration.\\
Since the neutron star viscosity is too low to lead to a tidal locking of the
system during the short interaction time, this spin configuration is not a 
realistic one (Bildsten and Cutler 1992). 
It is, however, straightforward to construct such {\em corotating equilibrium 
initial configurations}, a fact that makes these conditions appealing.
We use the full 3D-code to construct the initial particle distributions.
This is done in the corotating frame as here all the fluid velocities vanish.
The relaxation is performed in the following steps:
\begin{itemize}
\item calculate the gravitational and hydrodynamic forces,
$\vec{f}_{g,i}$ and $\vec{f}_{h,i}$, on each particle $i$ using the full hydro-code

\item determine the resulting force on the centre of mass of the neutron star
\begin{equation}
\vec{f}_{cm,ns}= \frac{\sum_i m_i( \vec{f}_{g,i} + \vec{f}_{h,i})}{M_{ns}}.
\end{equation}
Since our neutron star is located on the positive x-axis (origin at the centre of mass),
this force will point in negative x-direction.

\item In order to obtain a circular orbit, this force on the centre of 
mass has to be balanced exactly by the centrifugal forces on the centre 
of mass, which yields a condition for the rotation frequency:
\begin{equation}
\omega= \sqrt{\frac{|\vec{f}_{cm,ns}|}{M_{ns} a_{ns}}}, \label{omega_orb}
\end{equation}
where $a_{ns}$ is the distance of the neutron star from the origin.

\item Now the non-inertial forces are added:
\begin{equation}
\vec{f}_i= \vec{f}_{g,i} + \vec{f}_{h,i} - \vec{\omega} \times 
(\vec{\omega} \times \vec{r}_i), 
\end{equation}
where $\vec{r}_i$ is the position vector of particle $i$. Note that we have
omitted the Coriolis-force since it does not contribute to find the final
corotating state.
The addition of the Coriolis force during the relaxation process would only
introduce an additional lateral oscillation and therefore increase the 
computing time to find the equilibrium state.

\item We finally apply an additional velocity-proportional friction term
to force the system into the corotating state:
\begin{equation}
\vec{f}_i= \vec{f}_{g,i} + \vec{f}_{h,i} - \vec{\omega} \times 
(\vec{\omega} \times \vec{r}_i) - \frac{\vec{v}_i}{\tau},
\end{equation}
where $\tau$ is a damping time scale.

\item Once equilibrium is achieved the velocities are transformed back to 
the space fixed system and the radial velocities are added.
\end{itemize}

\begin{figure}
     \hspace*{0cm}\psfig{file=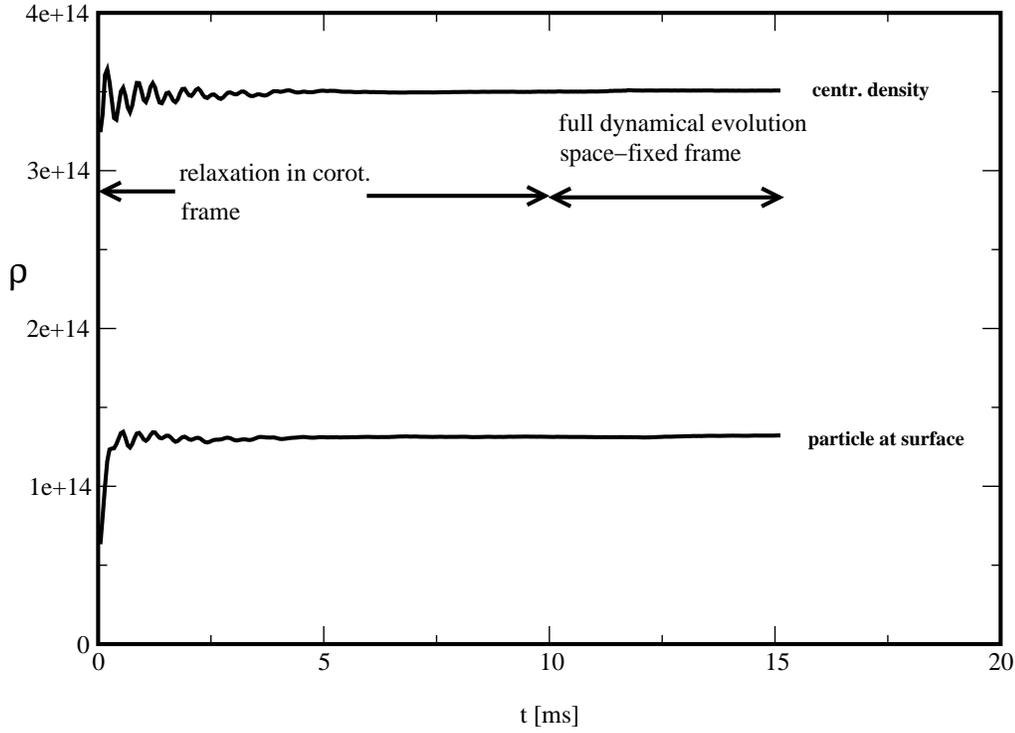,width=12cm,angle=-90}
    \caption{\label{cor_relax} Demonstration of the accuracy of our
corotating initial conditions.
During the first 10 ms the system is damped (in the corotating frame) into
equilibrium, then it is transformed to the space-fixed frame. Subsequently
a full simulation of one orbit is performed. The system remains perfectly in
the desired equilibrium.}
\end{figure}

An example of such a relaxation process for a 1.4 M$_{\odot}$ neutron star
and a 2.8 M$_{\odot}$ black hole is
shown in Fig. \ref{cor_relax}. In the first 10 ms the system is relaxed 
in the corotating frame. Then it is set to an orbit without radial velocities
and no backreaction forces are applied. Subsequently it is evolved for 
more than one orbital revolution with the full hydro code. The system 
remains in perfect equilibrium: no oscillation in the densities of the 
centre and the surface can be seen, after one orbit practically each 
particle lies right on top of its initial position.\\
For the {\em case without initial neutron star spin} we use a 1.4 \msun\ ns
that has been relaxed to equilibrium in isolation. This star is set (with the
appropriate radial velocity) on an orbit with an angular frequency
determined from the force balance condition, eq. (\ref{omega_orb}).

\section{Results}
%text plus runs + testruns + table
%
% - difficulty to form accretion disk
% - mini-ns
% - short description of evolution/morphology
%
%
% coldens-plots: i)   q=0.1, cor (6 panels)
%                ii)  ``irr''    (6 panels)
%                iii) q=0.93, cor(6 panels)
\begin{table*}
\caption{Summary of the different runs. C: corotation, I: no spin,
$a_0:$ initial separation; $T_{sim}:$ simulated duration; $\#$ part.: 
total particle number}
\begin{flushleft}
\begin{tabular}{ccccrccccc} \hline \noalign{\smallskip}
run & mass ratio q & spin &\# part. & $a_{0}$ [km] & $T_{sim}$ [ms] & remark\\ \hline \\
A   & 0.1  &   C           & 292 427 & 105 &  63.8 &\\ 
B   & 0.3  &   C           & 570 587 &  72 &  61.0 &\\
C   & 0.5  &   C           & 292 427 &  60 &  55.6 &\\
D   & 0.1  &   I           &1 083 218&  102&  21.9 &\\         
E   & 0.3  &   I           &1 083 218&  69 &  26.5 &\\
F   & 0.5  &   I           & 292 427 & 43.5&  24.7 &\\
    &      &               &         &    &        &\\
G   & 0.93 &   C           & 102 831 & 48 &  39.5  & extreme mass ratio\\
H   & 0.1  &   I           &  35513  & 100&  20.2  & polytrope, $\Gamma=2$\\
I   & 0.1  &   I           &  35513  & 100&  20.2  & polytrope, $\Gamma=3$\\
%J   & 0.08 &   C           &         &    &        & ns with 0.4 \msun\\

\end{tabular}
\end{flushleft}
\label{runs}
\end{table*} 
We performed six production runs with up to $10^{6}$ SPH-particles, exploring
the dependence of the results on the mass ratio ($q=0.1,0.3,0.5$) and on the 
initial neutron star spin (corotation and no initial spin). We follow the 
evolution for a long time (up to 64 ms corresponding to 300 dynamical time 
scales of the initial neutron star; previous investigations
(Lee and Kluzniak 1999a, 1999b, Lee 2000, Lee 2001, Janka et al. 1999)
typically simulated $\sim$ 20 ms). 
Details of the runs are shown in Table \ref{runs}.\\
After an inspiral of typically one orbit mass transfer sets in. In the
corotating case, the mass from the neutron star is initially fed
directly into the hole with mass transfer rates of several hundred \msun\ 
per second, see below. In agreement with the analytical estimates, the
system without initial spin has less difficulties to bypass the
Schwarzschild-radius. Although somewhat easier in the irrotational case, we
find it generally hard to form a massive accretion disk around the hole.\\
The dynamical evolution of the runs A (corotation, q= 0.1), run E (no spin, q=
0.3) and the extreme case run G (corotation, q= 0.93) are shown in
Figs. \ref{pmNq01cor}, \ref{pmNq03irr} and \ref{pmNq093} (colour coded is the 
logarithm of the column density in g/cm$^2$).

\subsection{Black Hole Evolution}
%
% -M_BH(t)/M_dot_BH
We plot in Fig. \ref{M_BH_cor} the evolution of the mass transfer rates 
{\em into the hole} (in solar masses per second) and the evolution of the
black hole masses for the corotating runs (runs A - C). For the smallest 
mass ratio, q= 0.1, there is one extreme initial mass transfer phase (see 
panel two in Figure \ref{pmNq01cor}), in which around 1 \msun\ is
transferred within 5 ms into the hole, corresponding to a peak rate of more
than 500 \msun /s. After this phase the system undergoes six more mass
transfer episodes in which the individual peak rates reach $\sim 0.8$ \msun
/s; the hole mass increases only insignificantly during this phase. The 
larger mass ratio cases exhibit a shorter initial mass transfer, however 
substantially more mass is accreted by the hole in the subsequent mass 
transfer episodes (see step-like behaviour in panel 2).\\
The mass transfer in the cases where the neutron star has no initial spin
(see Fig. \ref{M_BH_irr}) is also dominated by the initial mass transfer 
episode with huge rates, again up to 500 \msun /s. The rates in the 
subsequent phases, however, are typically two orders of magnitude lower 
so that the hole mass increases only marginally.

\subsection{Disk}
\begin{figure*}
  \begin{minipage}[t]{\columnwidth}
    \hspace*{0cm}\psfig{file=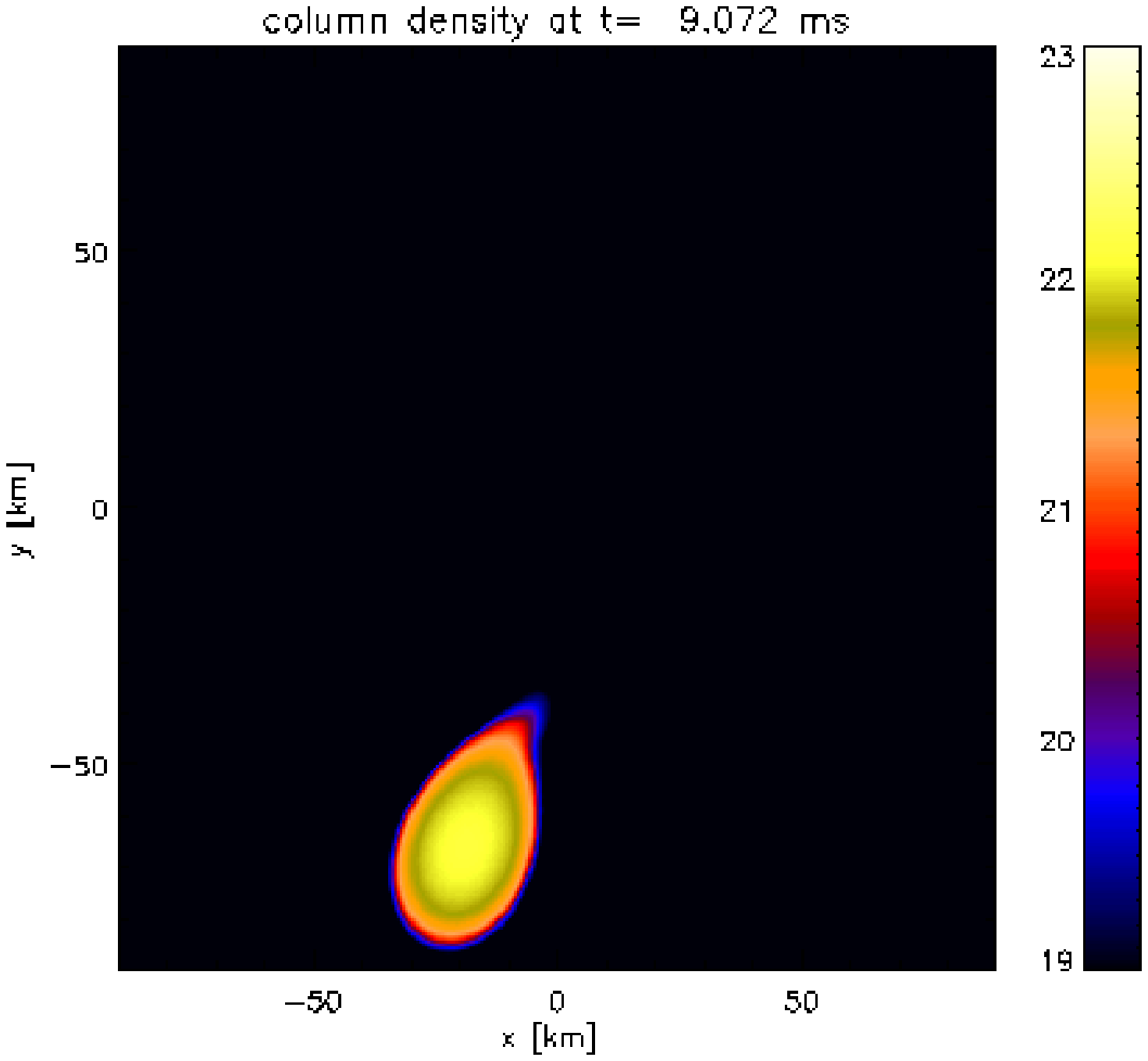,width=7.7cm,angle=0}\\
    \vspace*{1cm}
  \end{minipage}
  \hspace*{.6cm}
  \begin{minipage}[t]{\columnwidth}
    \psfig{file=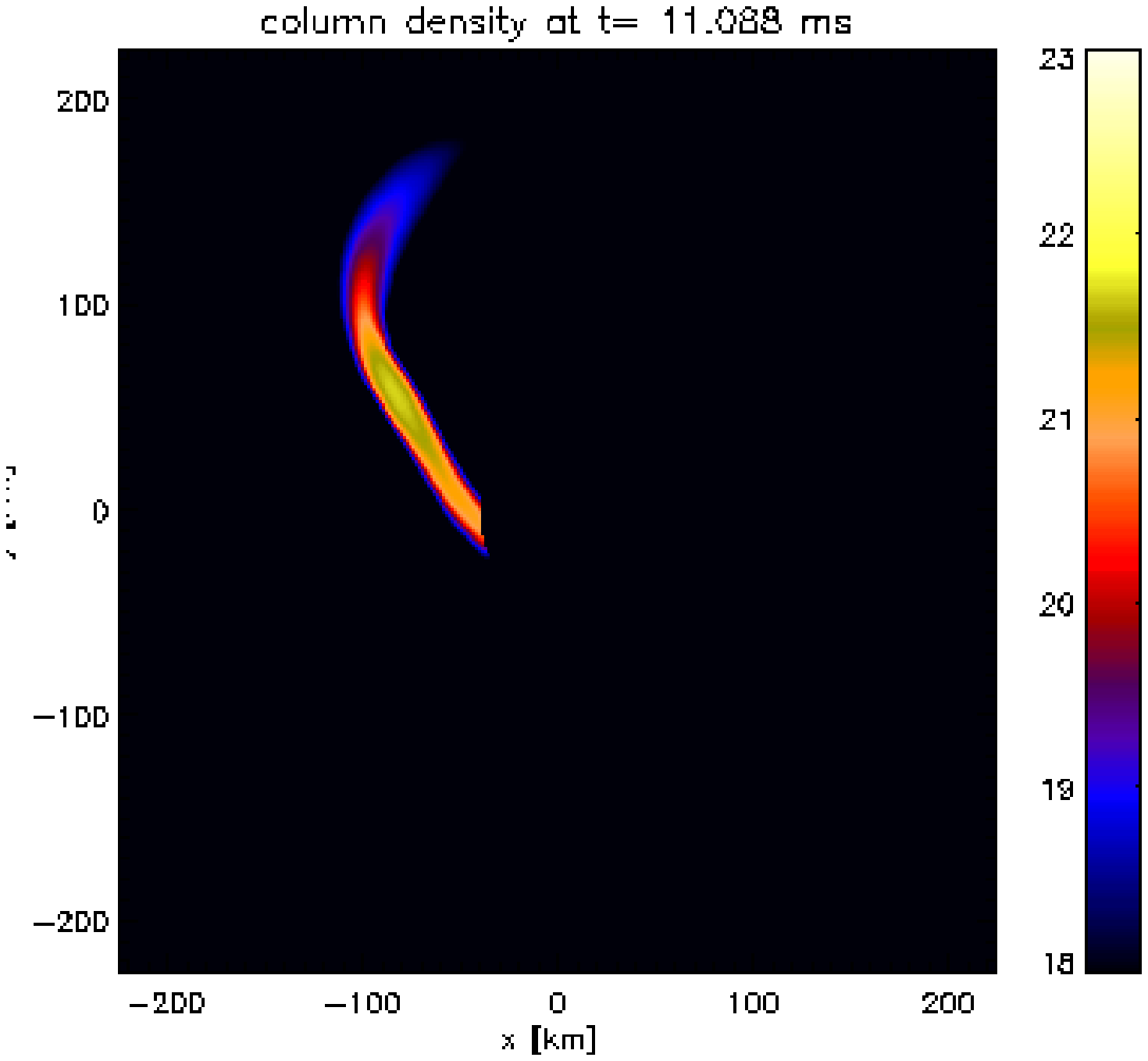,width=7.7cm,angle=0}
    \vspace*{1cm}
  \end{minipage}

\vspace*{-2cm}

  \begin{minipage}[]{\columnwidth}
    \hspace*{0cm}\psfig{file=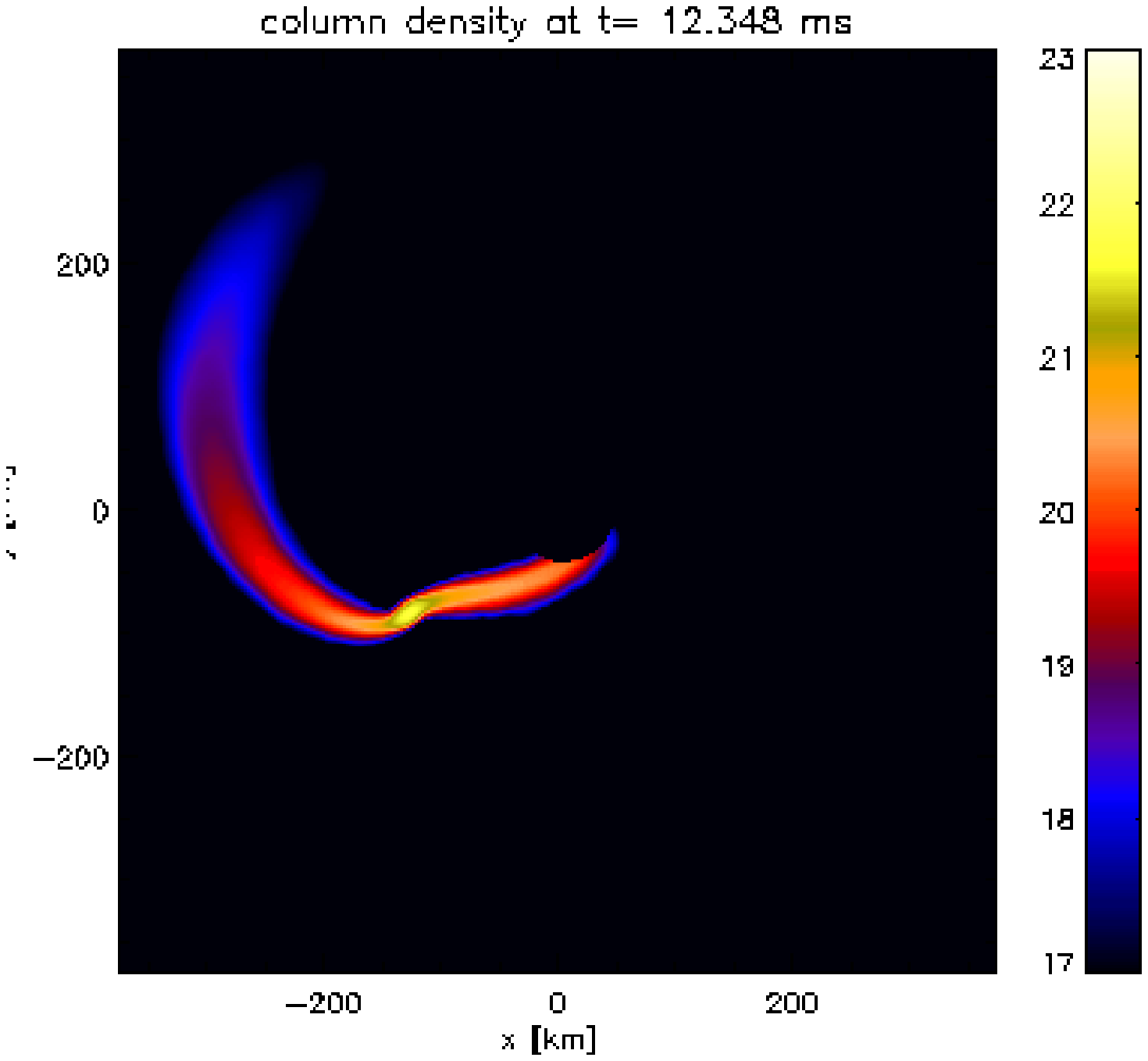,width=7.7cm,angle=0}\\
    \vspace*{1cm}
  \end{minipage}
  \hspace*{.6cm}
  \begin{minipage}[]{\columnwidth}
     \psfig{file=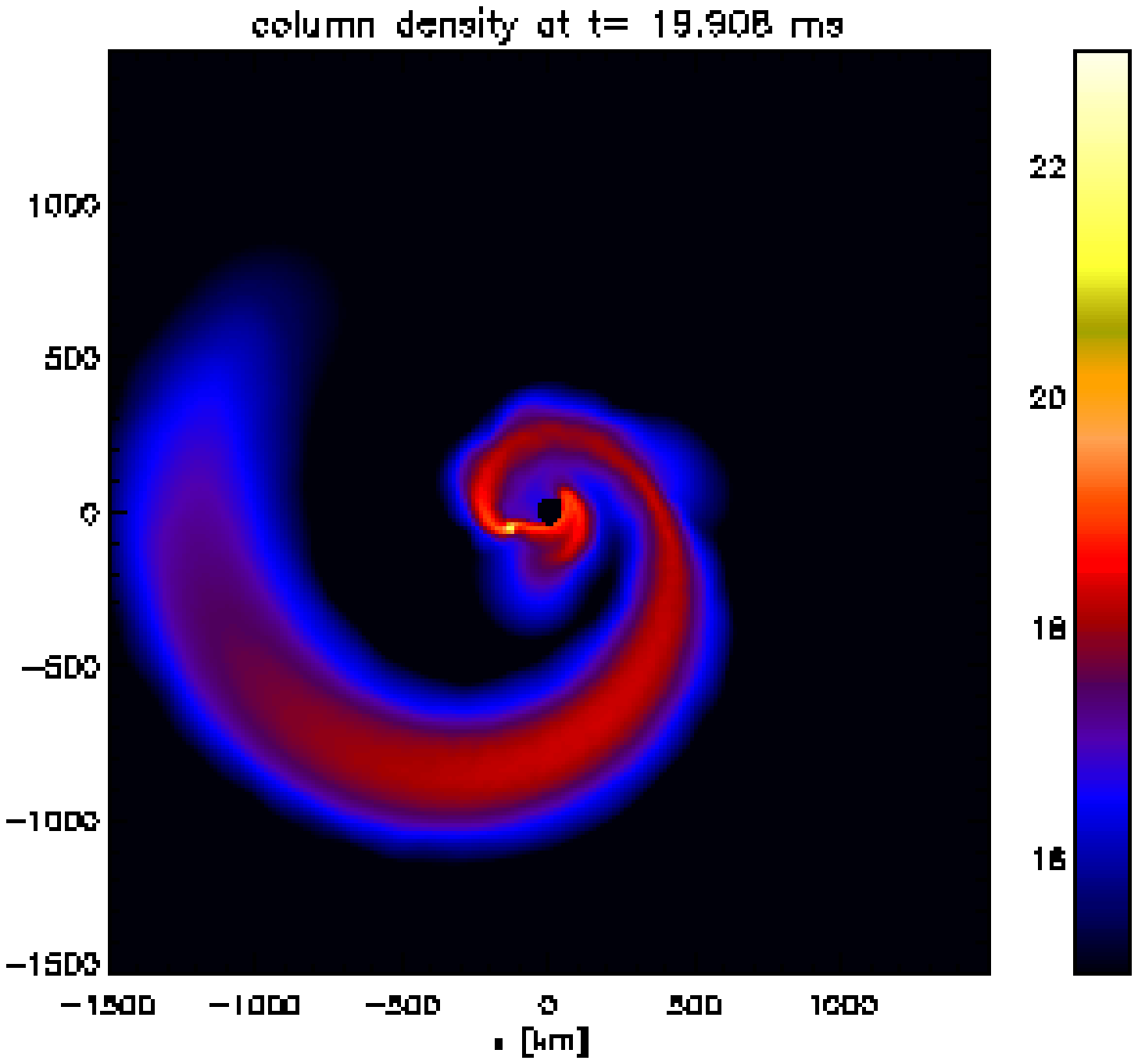,width=7.7cm,angle=0}
     \vspace*{1cm}
  \end{minipage}

\vspace*{-2cm}

   \begin{minipage}[]{\columnwidth}
     \hspace*{-0.4cm}\psfig{file=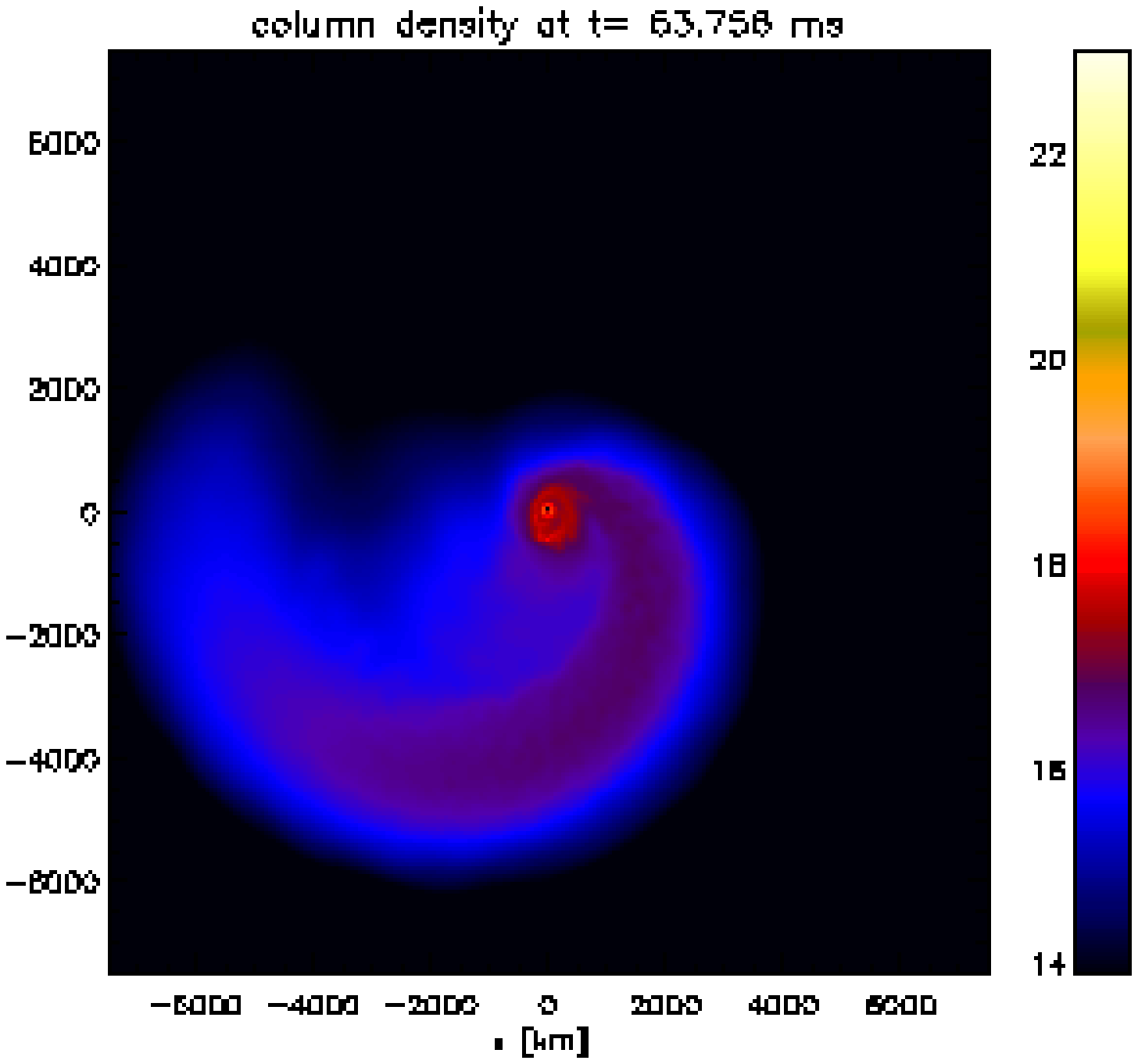,width=7.7cm,angle=0}\\
     \vspace*{1cm}
   \end{minipage}
   \begin{minipage}[]{\columnwidth}
     \vspace*{-1cm}
 \hspace*{0.3cm}\psfig{file=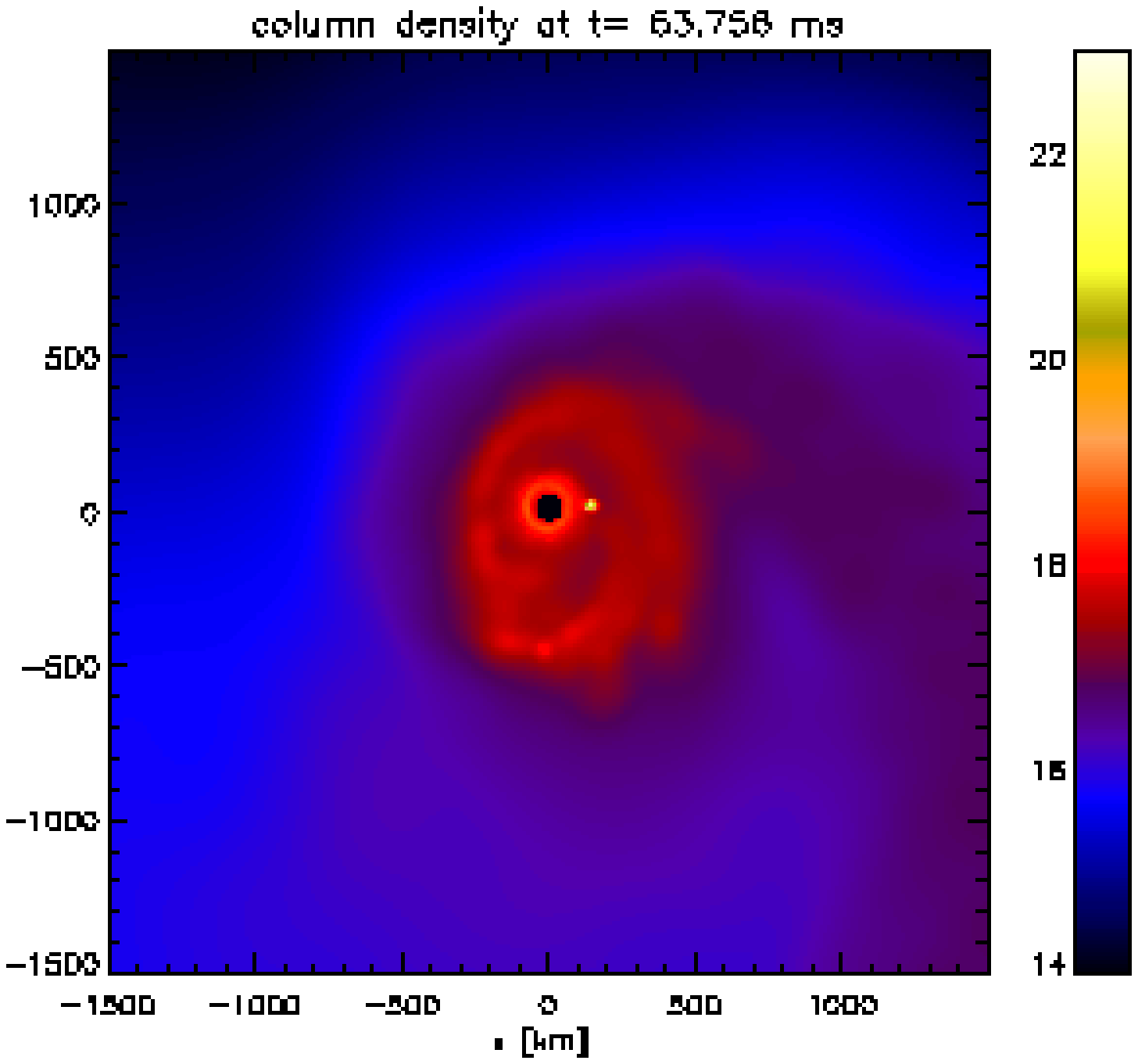,width=7.7cm,angle=0}\\
   \end{minipage}

    \caption{Dynamical evolution of run A (corotating neutron star, mass ratio
    $q=0.1$). A ``mini neutron star'' survives the encounter and keeps
    orbiting the central black hole. At the moment shown in panel six (=
    blow-up of panel five) the mini neutron star has completed seven orbital
    revolutions. Colour-coded is column density.
\label{pmNq01cor}}
\end{figure*}

\begin{figure*}
  \begin{minipage}[t]{\columnwidth}
    \hspace*{0cm}\psfig{file=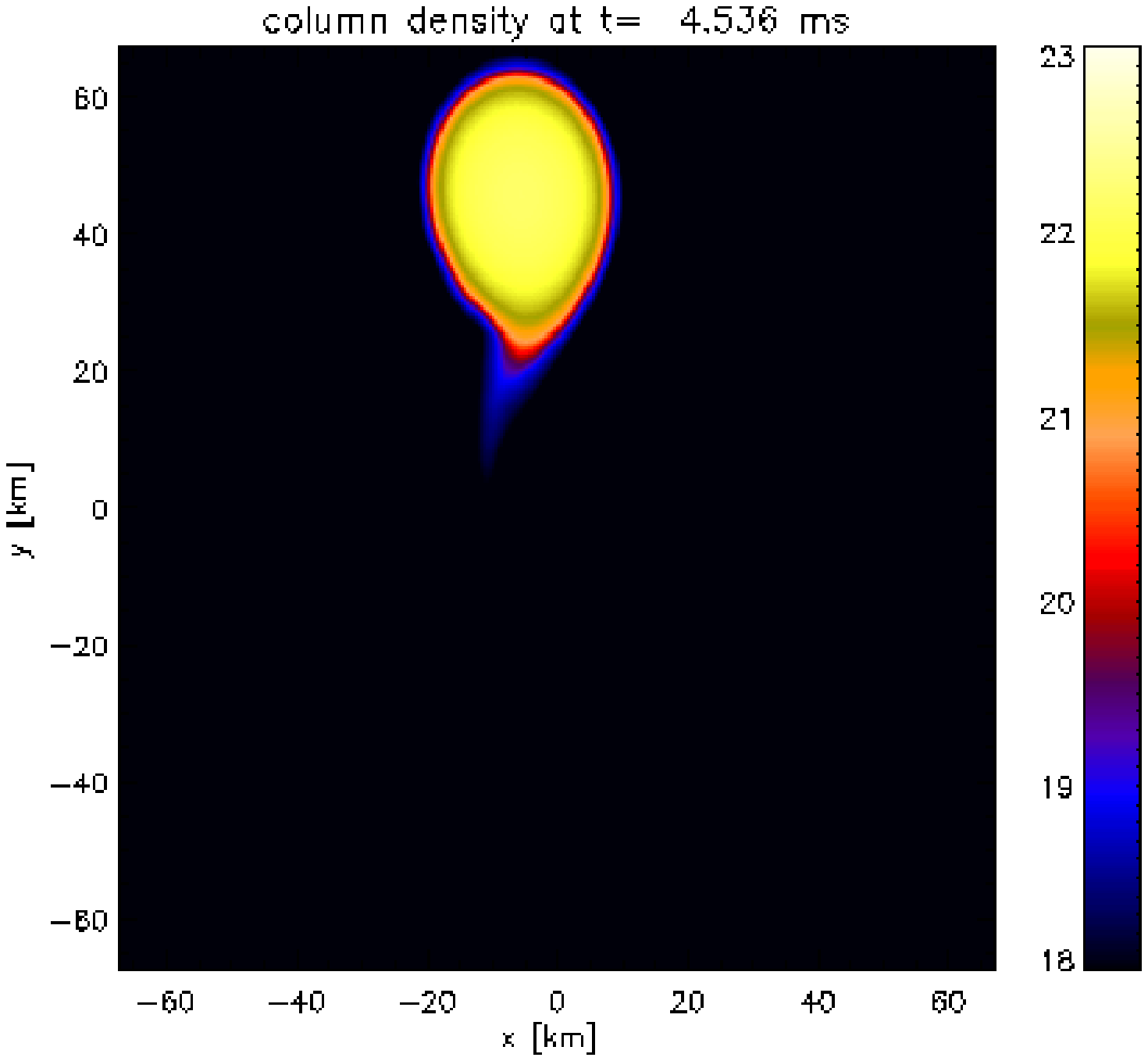,width=7.7cm,angle=0}\\
    \vspace*{1cm}
  \end{minipage}
  \hspace*{.6cm}
  \begin{minipage}[t]{\columnwidth}
    \psfig{file=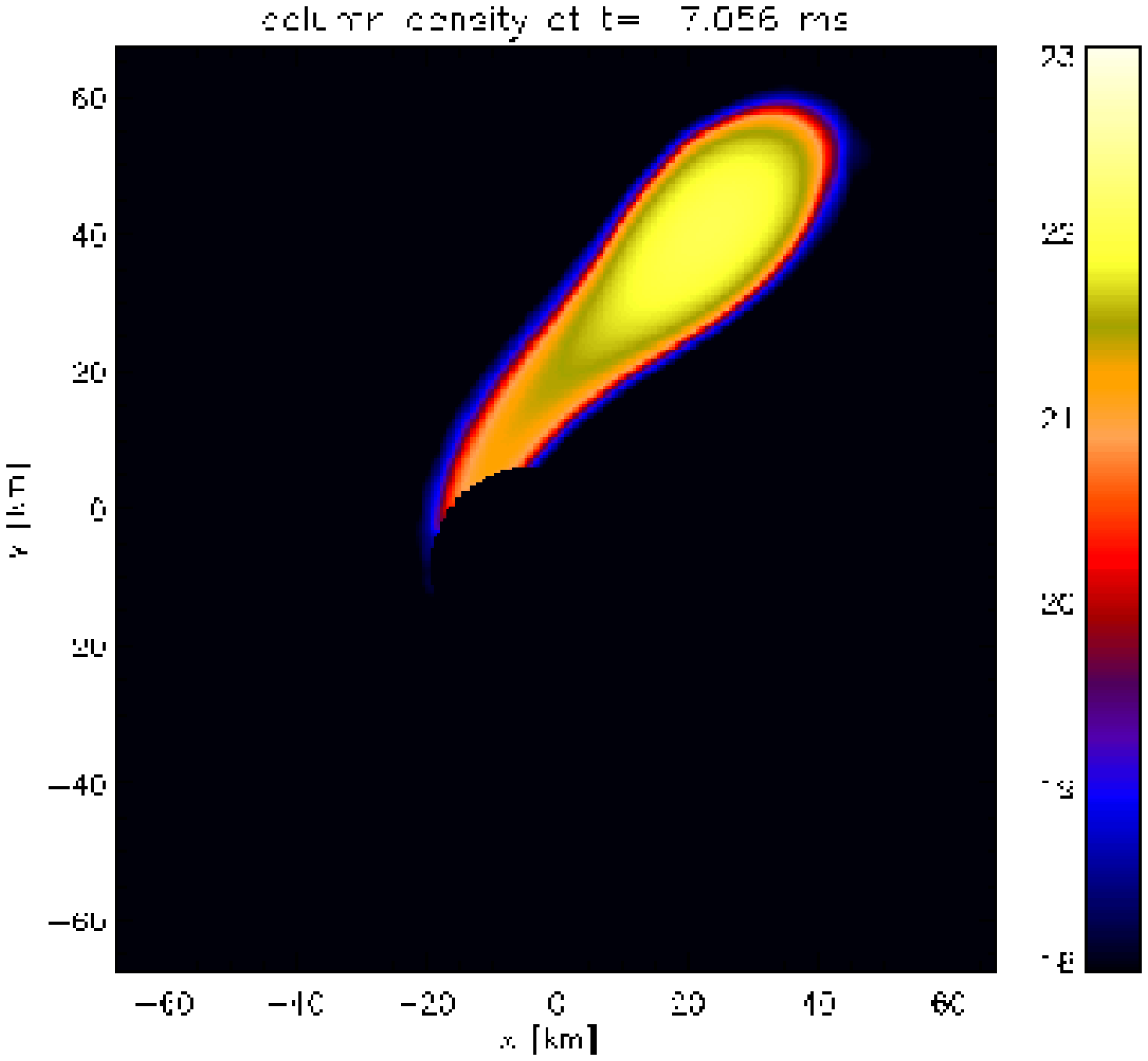,width=7.7cm,angle=0}
    \vspace*{1cm}
  \end{minipage}

\vspace*{-2cm}

  \begin{minipage}[]{\columnwidth}
    \hspace*{0cm}\psfig{file=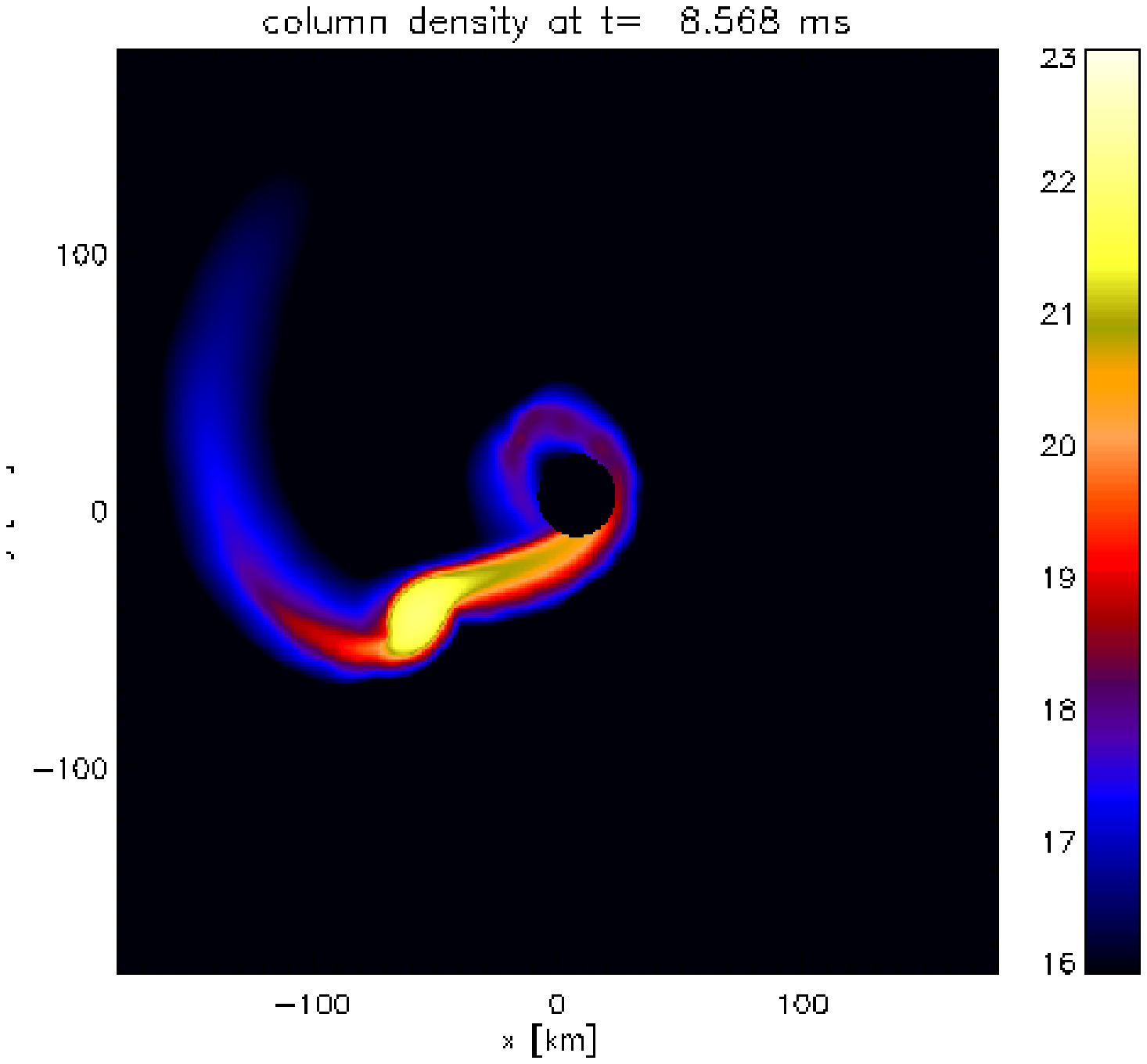,width=7.7cm,angle=0}\\
    \vspace*{1cm}
  \end{minipage}
  \hspace*{.6cm}
  \begin{minipage}[]{\columnwidth}
     \psfig{file=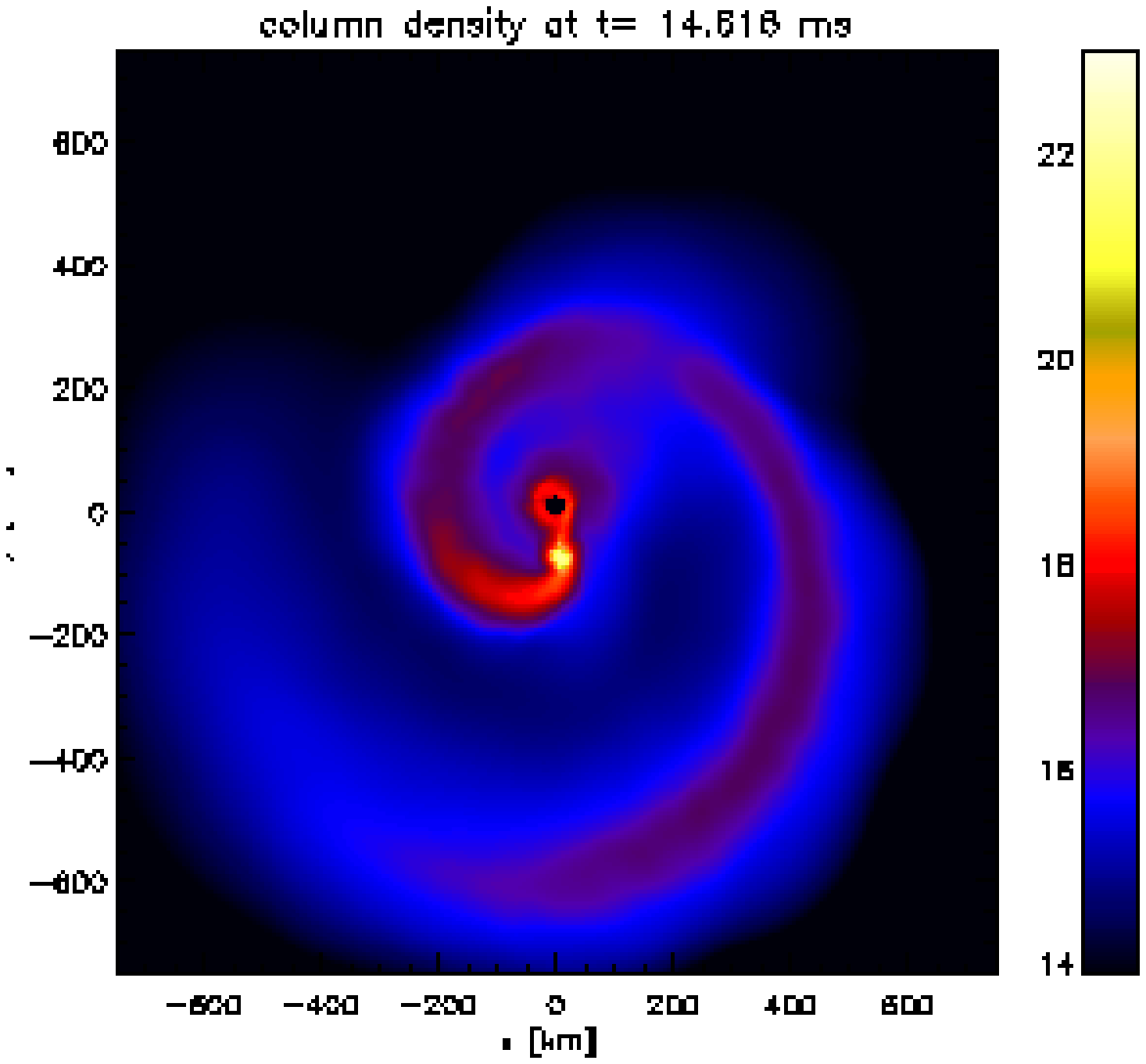,width=7.7cm,angle=0}
     \vspace*{1cm}
  \end{minipage}

\vspace*{-2cm}

   \begin{minipage}[]{\columnwidth}
     \hspace*{-0.4cm}\psfig{file=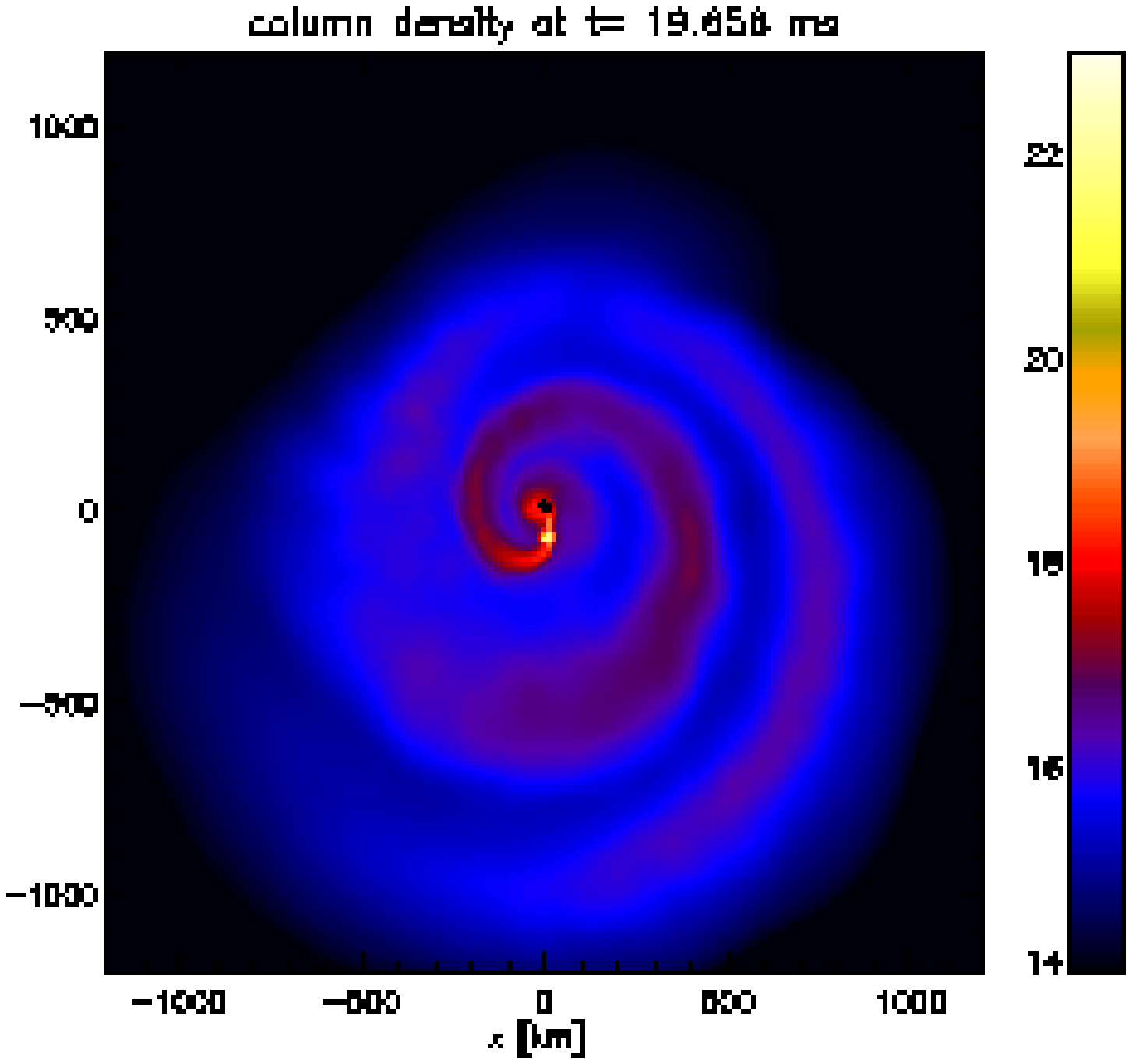 ,width=7.7cm,angle=0}\\
     \vspace*{1cm}
   \end{minipage}
   \begin{minipage}[]{\columnwidth}
     \vspace*{-1cm}
 \hspace*{0.3cm}\psfig{file=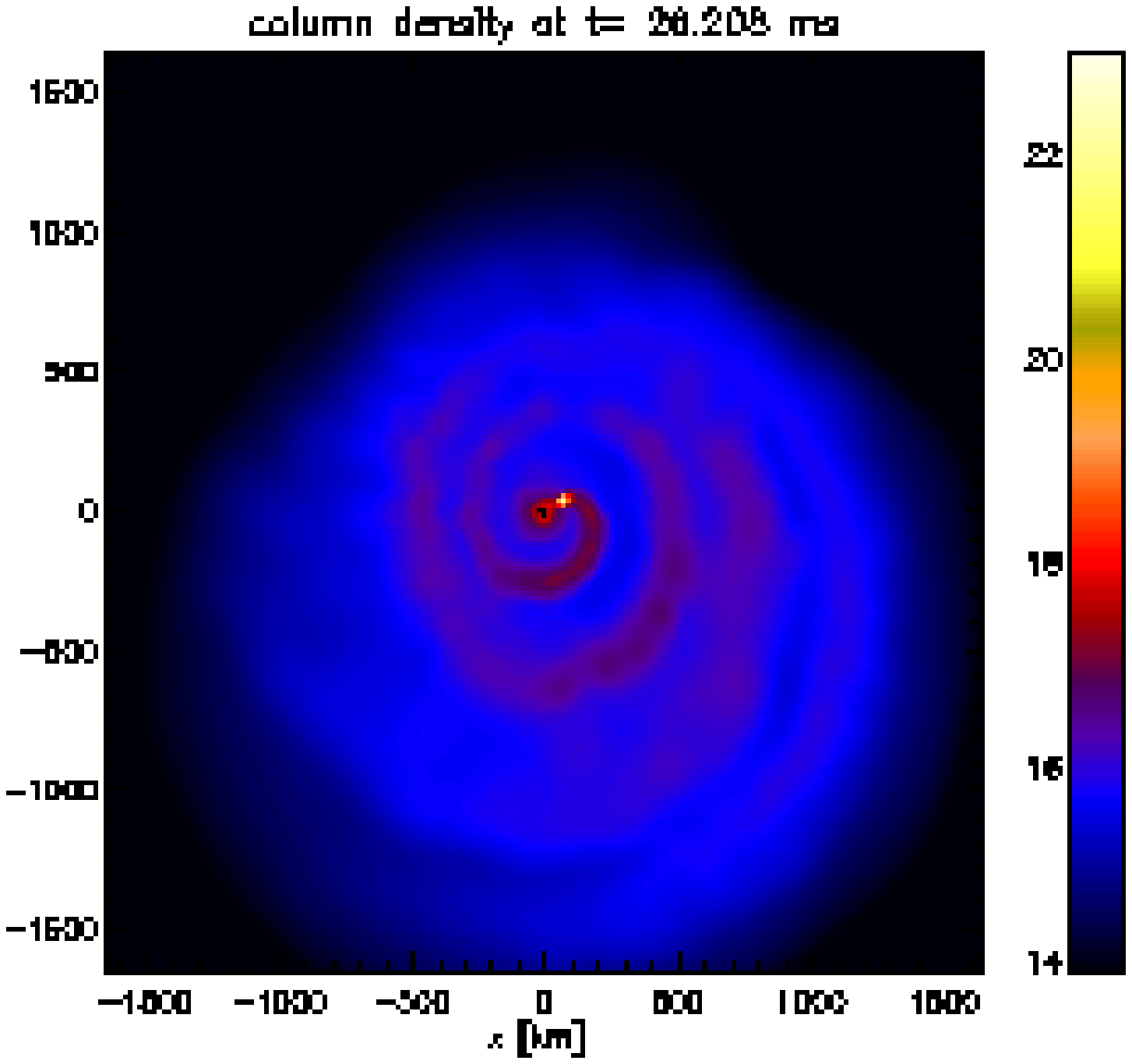,width=7.7cm,angle=0}\\
   \end{minipage}

    \caption{Evolution of run E (no neutron star spin, mass ratio q= 0.3).
\label{pmNq03irr}}
\end{figure*}

\begin{figure*}
  \begin{minipage}[t]{\columnwidth}
    \hspace*{0cm}\psfig{file=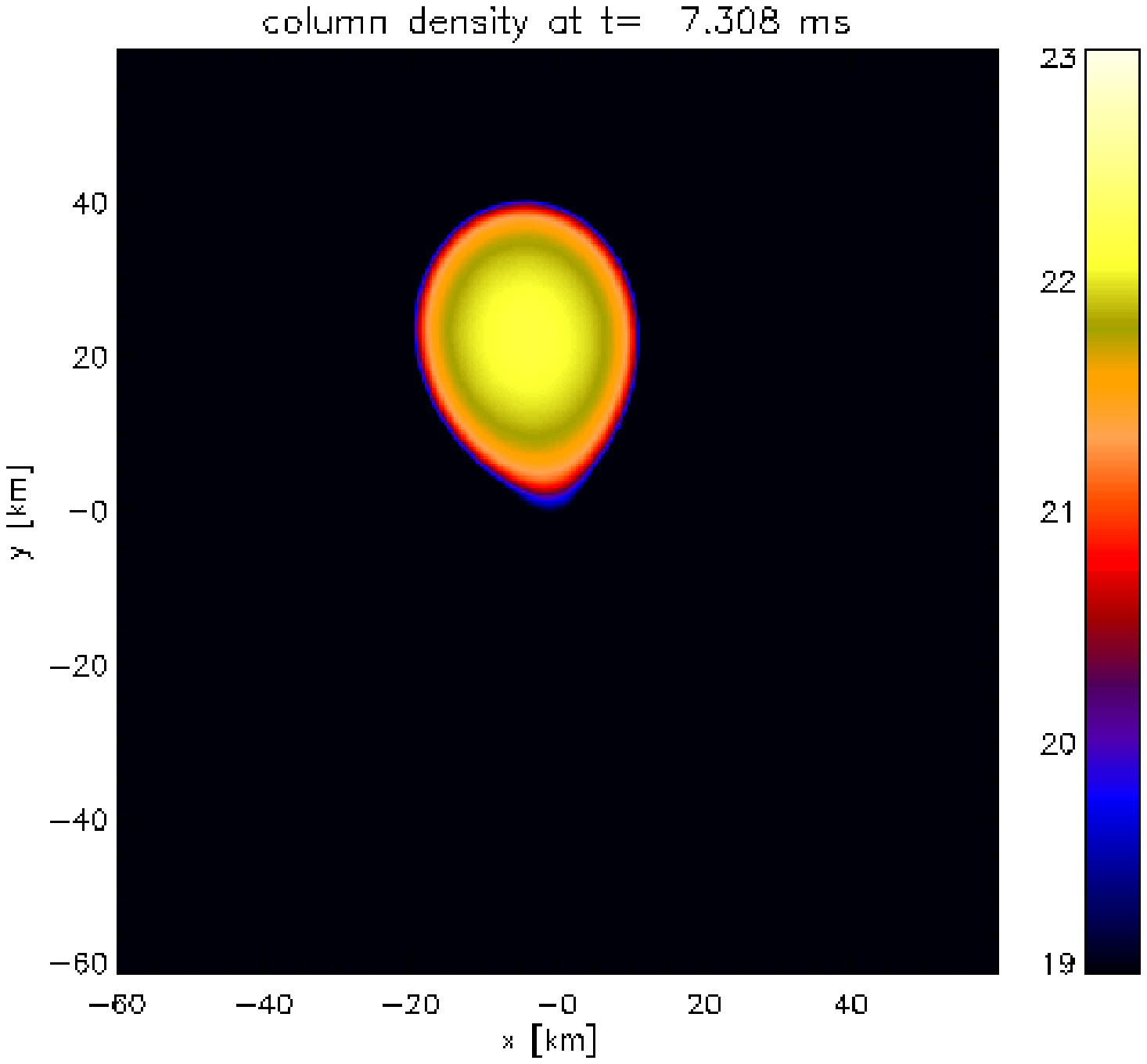,width=7.7cm,angle=0}\\
    \vspace*{1cm}
  \end{minipage}
  \hspace*{.6cm}
  \begin{minipage}[t]{\columnwidth}
    \psfig{file=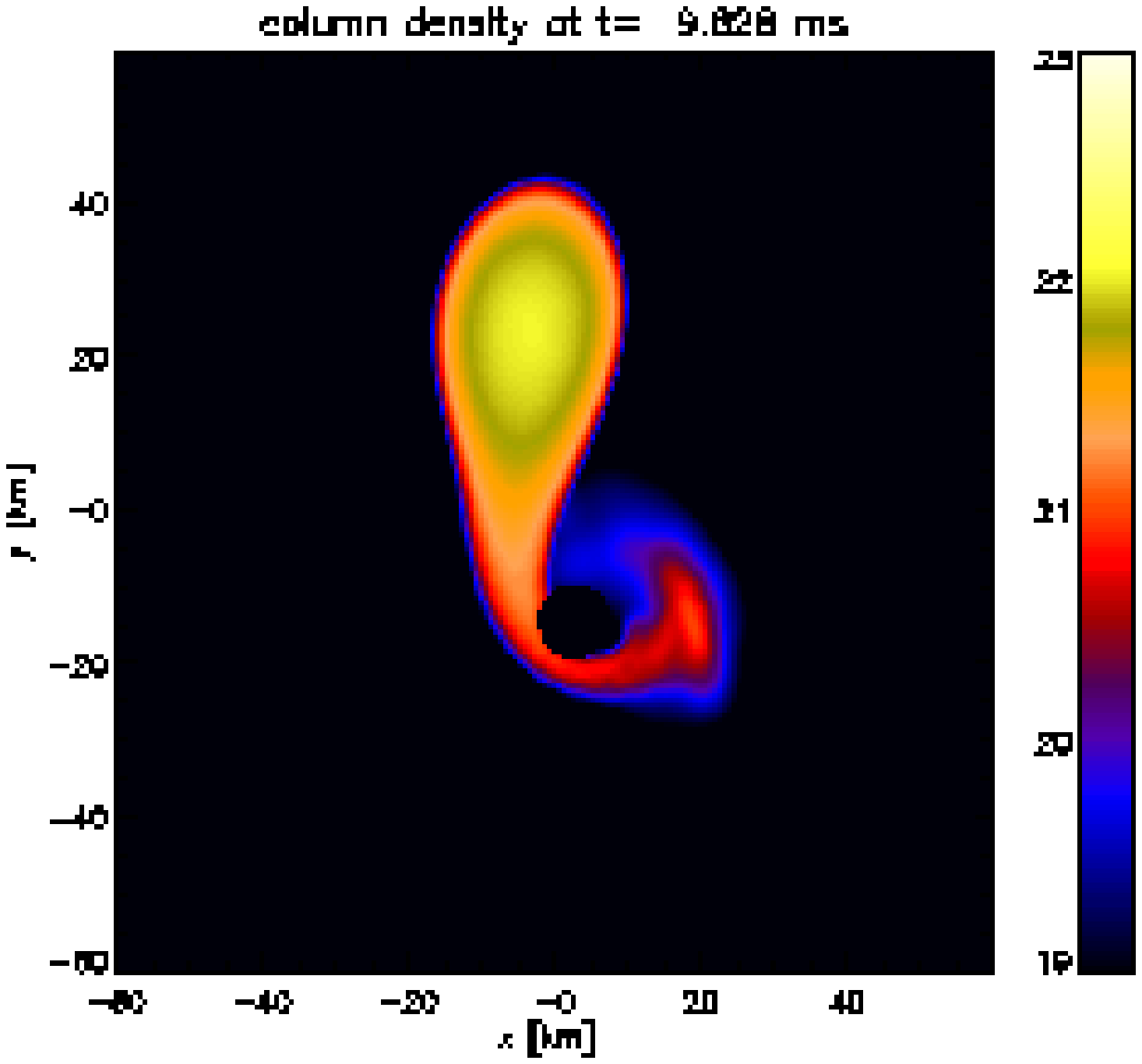,width=7.7cm,angle=0}
    \vspace*{1cm}
  \end{minipage}

\vspace*{-2cm}

  \begin{minipage}[]{\columnwidth}
    \hspace*{0cm}\psfig{file=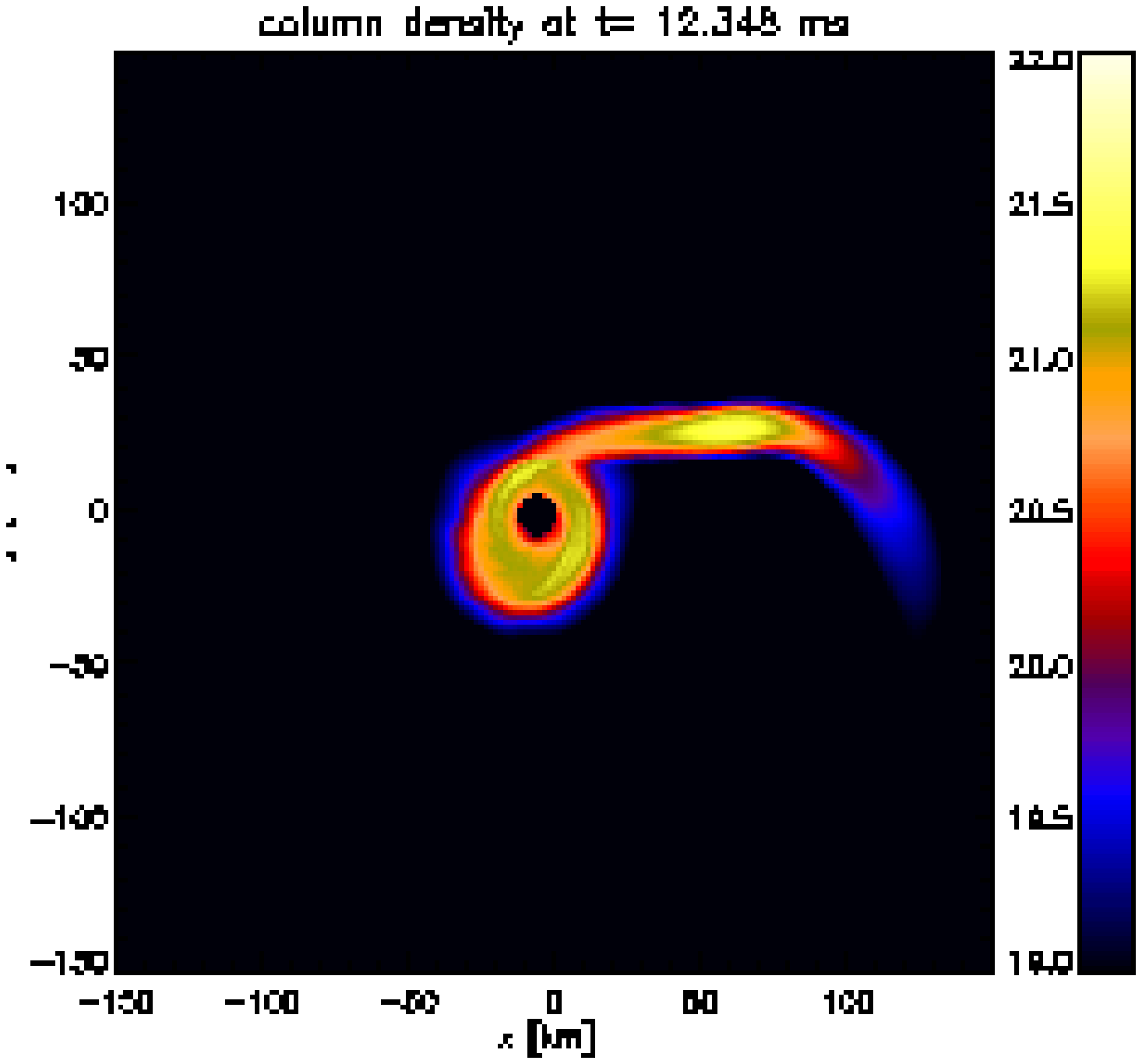,width=7.7cm,angle=0}\\
    \vspace*{1cm}
  \end{minipage}
  \hspace*{.6cm}
  \begin{minipage}[]{\columnwidth}
     \psfig{file=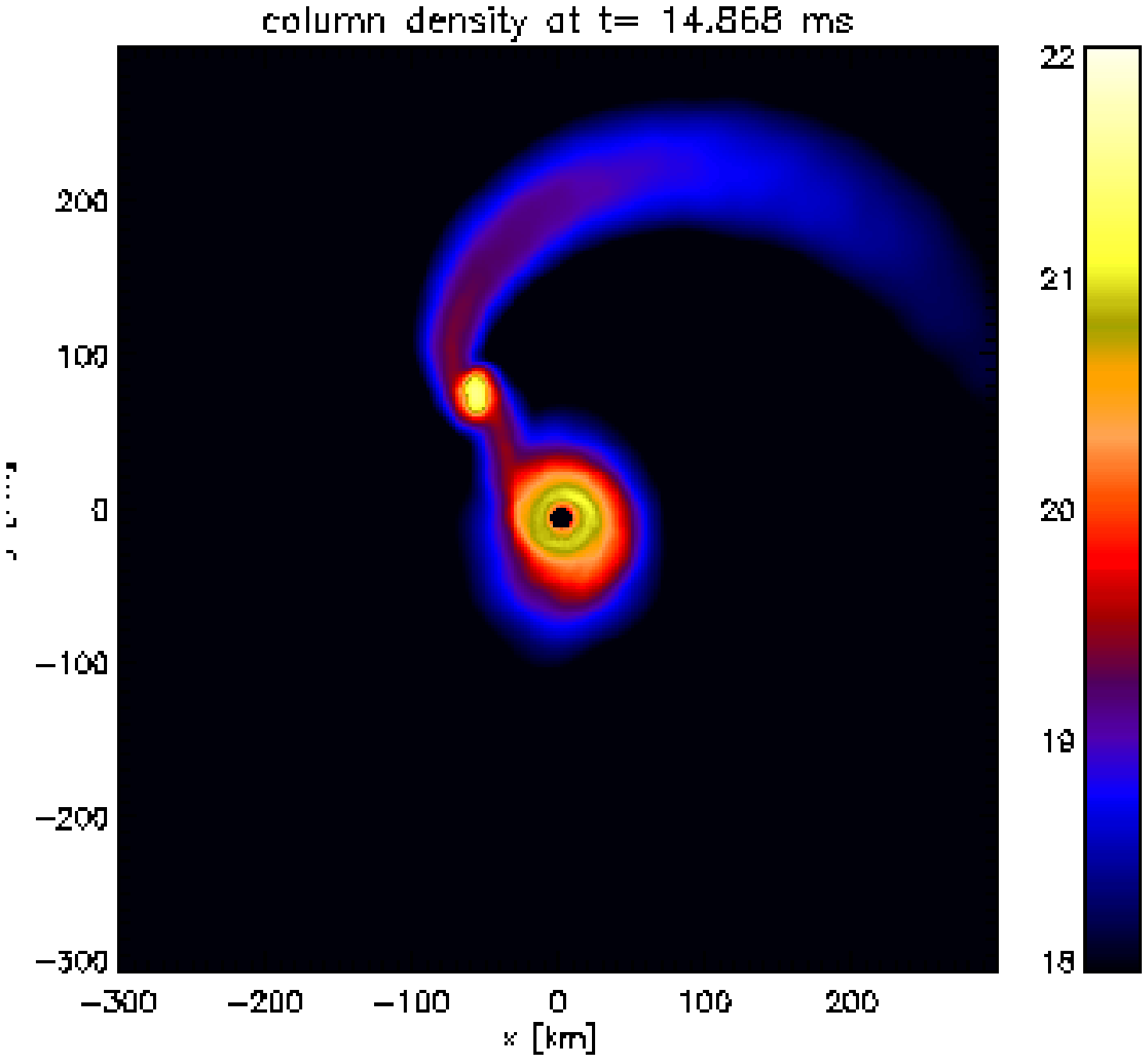,width=7.7cm,angle=0}
     \vspace*{1cm}
  \end{minipage}

\vspace*{-2cm}

   \begin{minipage}[]{\columnwidth}
     \hspace*{-0.4cm}\psfig{file=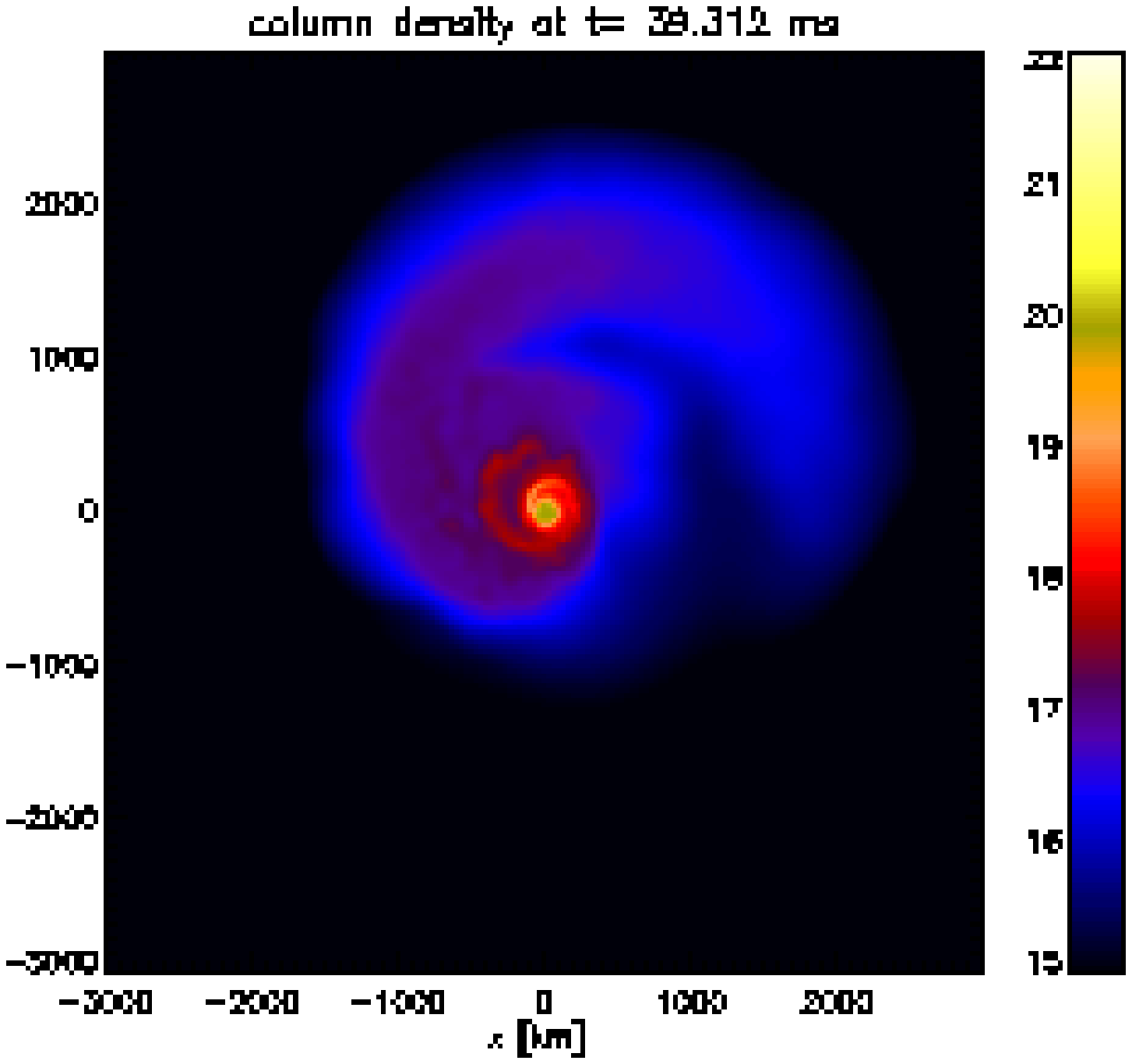,width=7.7cm,angle=0}\\
     \vspace*{1cm}
   \end{minipage}
   \begin{minipage}[]{\columnwidth}
     \vspace*{-1cm}
 \hspace*{0.3cm}\psfig{file=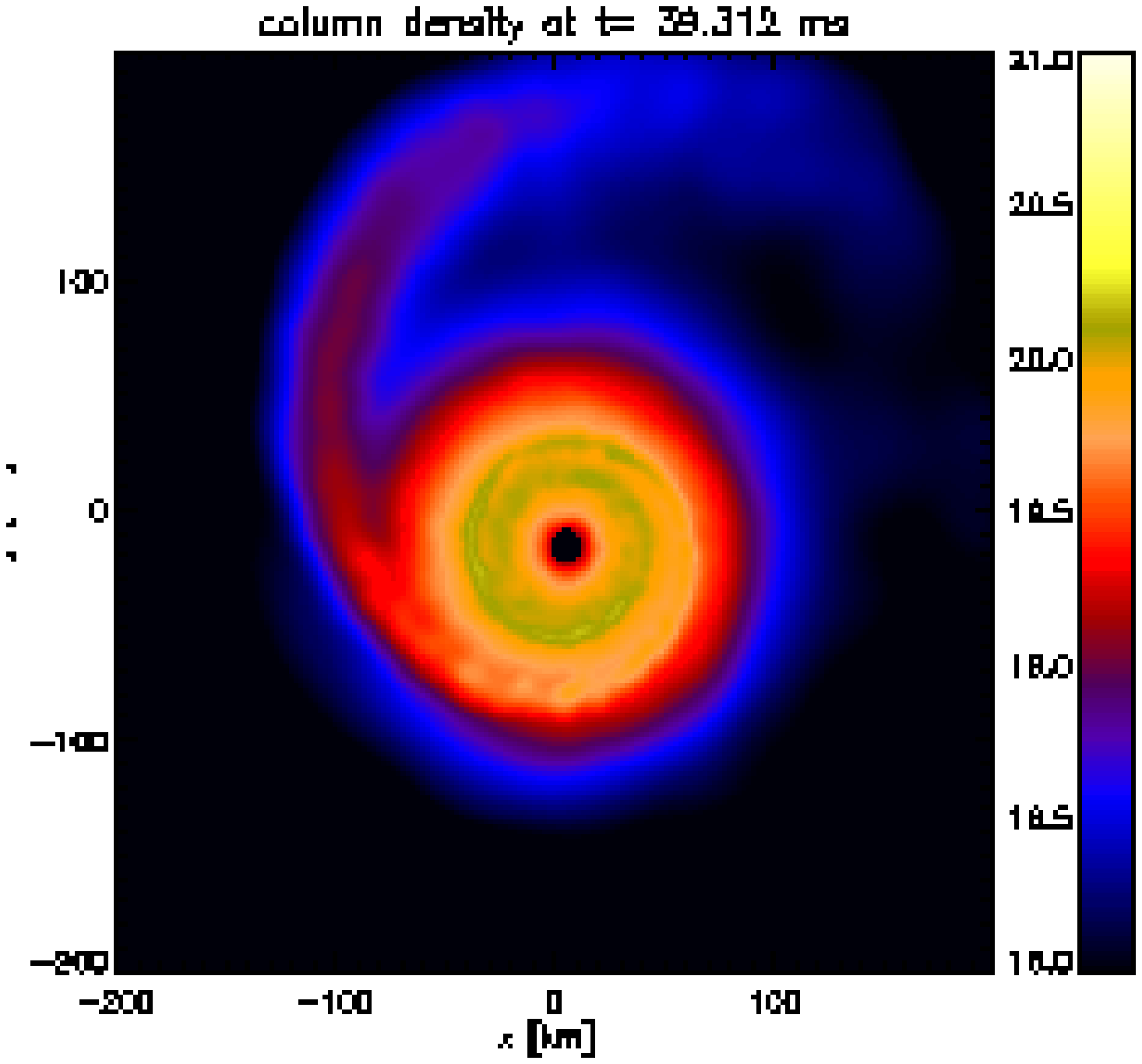,width=7.7cm,angle=0}\\
   \end{minipage}

    \caption{\label{pmNq093} Evolution of the extreme case, run G
    (corotation, black hole with 1.5 \msun\ and 1.4 \msun\ neutron star). It is
    only in this case that the neutron star is completely disrupted and a very
    massive ($\sim 0.3$ \msun) and hot (up to 10 MeV) accretion disk
    forms. Note that panel 6 shows a blow-up of the central region of panel five.}
\end{figure*}

\begin{figure}    
%\hspace*{-0.5cm}\psfig{file=Mdot_M_BH_cor.ps,width=8cm,angle=-90}
\psfig{file=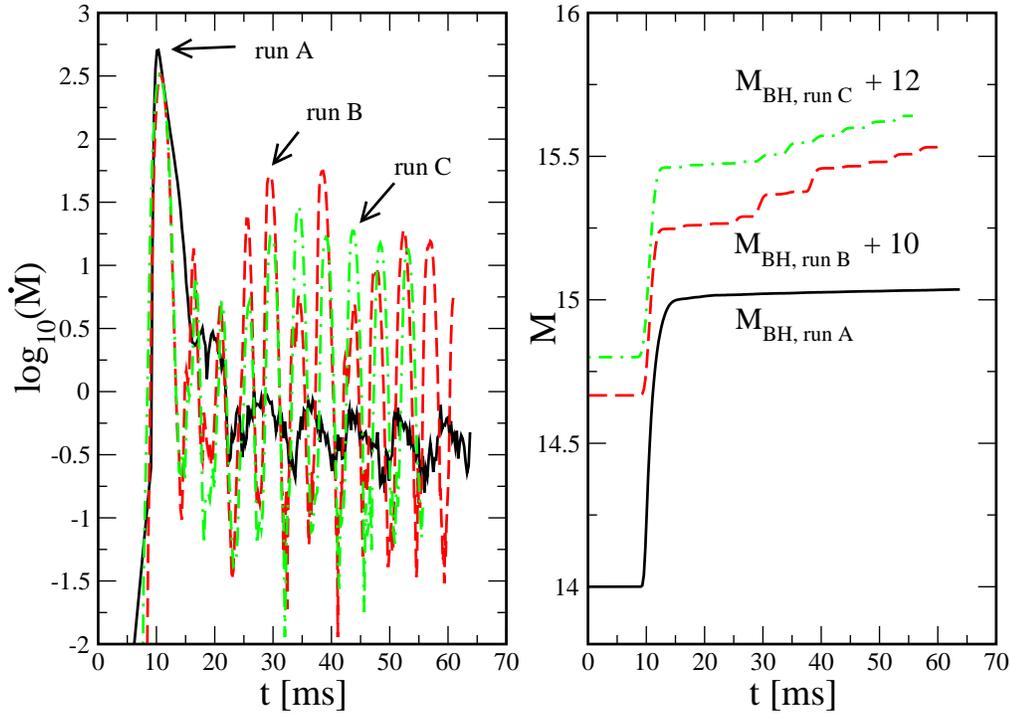,width=12cm,angle=-90}
    \caption{\label{M_BH_cor} Evolution of the black hole for the corotating
    cases. The left panel shows the logarithm of the mass transfer rate {\em
    into the hole} (in solar masses per second), $\dot{M}$. The right panel
    shows the evolution of the black hole mass (in solar masses). Note that
    suitable offsets have been added to enhance the visibility of the mass increase.}
\end{figure}

\begin{figure}    
\psfig{file=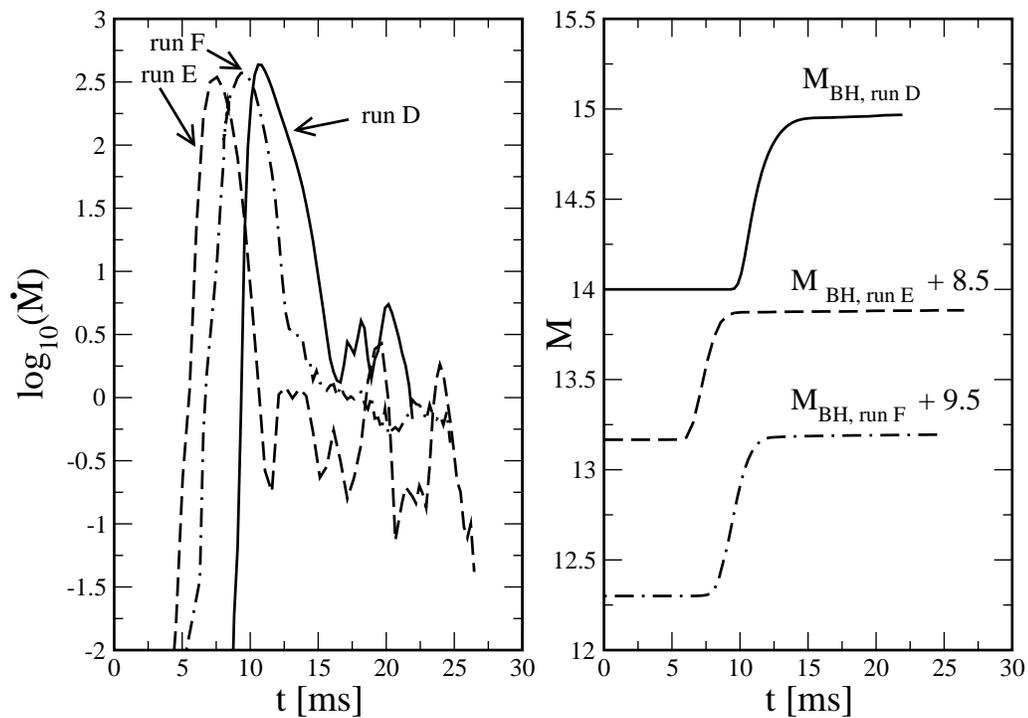,width=12cm,angle=-90}
    \caption{\label{M_BH_irr} Same as previous plot but for the cases without
    neutron star spin (masses in solar units; offsets added to enhance visibility).}
\end{figure}

\begin{figure}    
%\hspace*{0cm}\psfig{file=lrhoi_r_pmNq01cor.ps,width=7cm,angle=-90}
\hspace*{0cm}\psfig{file=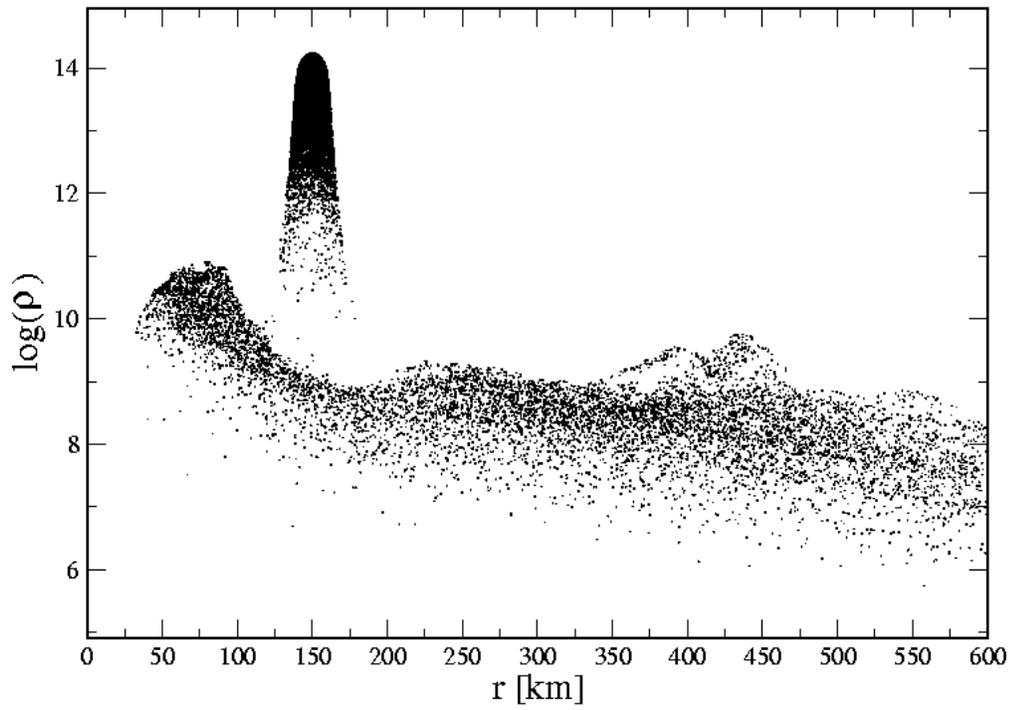,width=12cm,angle=-90}
\caption{\label{disk_runA} Logarithm of the densities of the SPH-particles in
  the inner region of run A (corresponding to the last panel in
  Fig. \ref{pmNq01cor}). Clearly visible is the ``mini neutron star'' at a
  radius of 150 km.}
\end{figure}

\begin{figure}    
\hspace*{0cm}\psfig{file=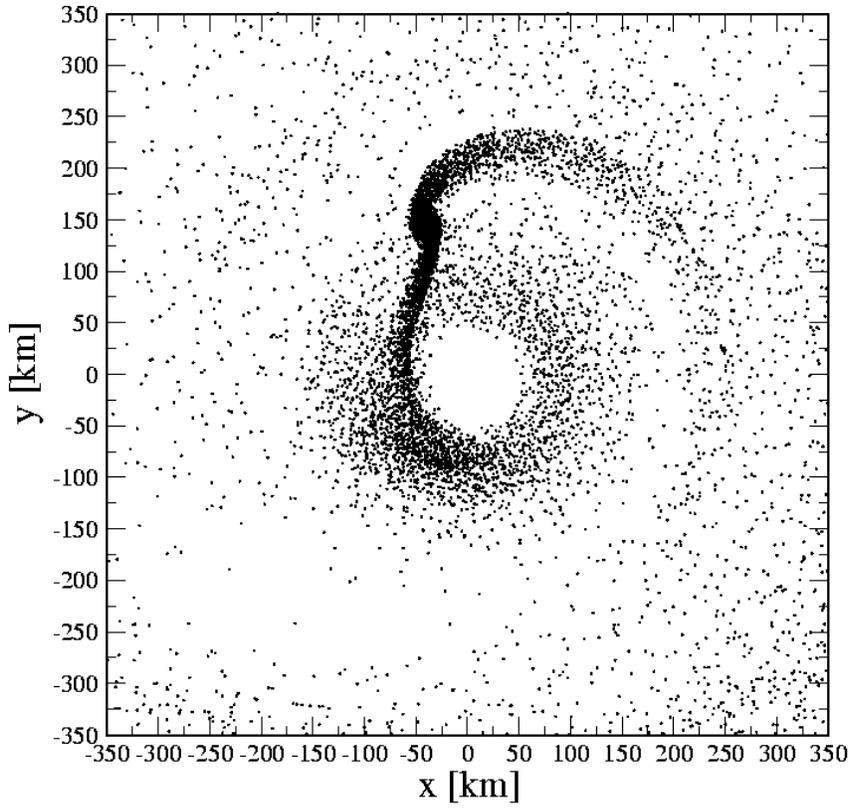,width=12cm,angle=-90}
\caption{\label{disk_stream} Interaction of the accretion stream with the
  disk (run A). Rather than a 'hot spot' like in a CV a 'hot band' accross the disk is
  visible.}
\end{figure}

% - comparison with analytical estimates
% - M_disk(t) (definition ?????)
% - ev. <T>_disk(t)
% - M_dot_disk(t)
An accretion disk around a black hole is a vital ingredient for virtually all
GRB-models. Usually it is assumed that a neutron star black hole merger is
just a variant of neutron star binary mergers and produces even more
massive (several 0.1 \msun), hot disks around the hole. We will show here that 
this assumption is -at least for our (hard) equation of state and the simulate
time scale of several 10 ms- not justified.\\
Fig. \ref{disk_runA} shows the SPH-particle densities in the inner region of
run A (compare to the last panel in Fig. \ref{pmNq01cor}). The innermost part
of the disk reaches densities of up to $10^{11}$ g cm$^{-3}$  and contains a
mass of only $7 \cdot 10^{-3}$ \msun. The disk out to 600 km (red area in
panel six of Fig. \ref{pmNq01cor}) contains $3 \cdot 10^{-2}$ \msun. 
The interaction of the accretion stream with the disk is shown in Fig. \ref{disk_stream}.
The mass-weighted temperatures of these disk regions
are 2.5 respectively 1.4 MeV. This configuration is not very promising to
launch a GRB, it can only anchor moderate magnetic fields and the neutrino
emission is considerably lower than in the double neutron star case which can
only produce GRB's of relatively low total energy output.\\
The other two corotating runs do not form any resolvable disk at all. The
matter from the neutron star is transferred directly into the hole. From its
back side mass is tidally removed. But rather than forming a massive tidal tail
like in the case of run A, this decompressed neutron star material forms a
low-density common envelope around the orbiting neutron star black hole
system.\\
The  case of q= 0.3  and no spin (run E) is shown in Fig. \ref{pmNq03irr}. Here
the rest of the neutron star is orbiting inside a disk that is continually
decreasing in density from values of $10^{10}$ g cm$^{-3}$ between the hole
and the neutron star to values of $\sim 10^5$ \gcc at distance of 400 km from
the hole. All the disk masses are very low, a few times $10^{-3}$ \msun, as 
still a lot of the total mass is retained in the orbiting mini-neutron star. 
The (mass averaged) temperatures in these inner disks are only moderate, 
typically 3 MeV. Further details can be found in Table \ref{final_config}.
Thus, none of the performed runs yields accretion disk that would be
convincing candidate to launch a GRB.

%
%--summarize some results in table
%
\begin{table*}
\caption{End of simulation: $M_{14}$: mass of orbiting mini-neutron star,
  $M_{disk}(T_{sim})$: disk mass, $\langle T\rangle _{disk}(T_{sim})$: mass
  averaged disk temperature, L$_{GW}^{max}$: peak gravitational wave
  luminosity, $\Delta E_{GW}$: energy lost in gravitational waves during the
  simulation time. In two of the cases no resolvable disks form.}
\begin{flushleft}
\begin{tabular}{cccccccccc} \hline \noalign{\smallskip}
run & $M_{14}(T_{sim})$ [\msun] & $M_{disk}(T_{sim})$ [\msun]& $\langle T
\rangle _{disk}(T_{sim})$ [MeV]& L$_{GW}^{max}$ [erg/s] & $\Delta E_{GW}$ [erg]  \\ \hline \\
A   & 0.15   & $6.2 \cdot 10^{-3}$ $^*$   & 2.5 & $5.9 \cdot 10^{55}$  & $1.3 \cdot 10^{53}$  & \\
B   & 0.40   & --& -- & $1.3 \cdot 10^{55}$  & $6.9 \cdot 10^{52}$  & \\
C   & 0.46   & --& -- & $3.7 \cdot 10^{54}$ & $3.8 \cdot 10^{52}$  & \\
D   & 0.24   & $ 4.9\cdot 10^{-3}$ & 2.7 & $4.7 \cdot 10^{55}$  & $1.7 \cdot 10^{53}$  & \\
E   & 0.68   & $2.3 \cdot 10^{-3}$ & 2.9 & $5.8 \cdot 10^{54}$  & $3.9 \cdot 10^{52}$  & \\
F   & 0.38   & $6.6 \cdot 10^{-3}$ & 3.9 & $6.5 \cdot 10^{54}$ & $3.1 \cdot 10^{52}$ & \\
\end{tabular}

\end{flushleft}
\label{final_config}
\begin{flushleft} $^*$ 'inner disk', see text \end{flushleft}
\end{table*} 

\subsection{Neutron Star Evolution}
\begin{figure}    
%\hspace*{0cm}\psfig{file=velocity_pmNq03cor_d35.ps,width=9cm,angle=0}
%\hspace*{0cm}\psfig{file=velocity_pmNq03ir_d11.ps,width=9cm,angle=0}
\hspace*{0cm}\psfig{file=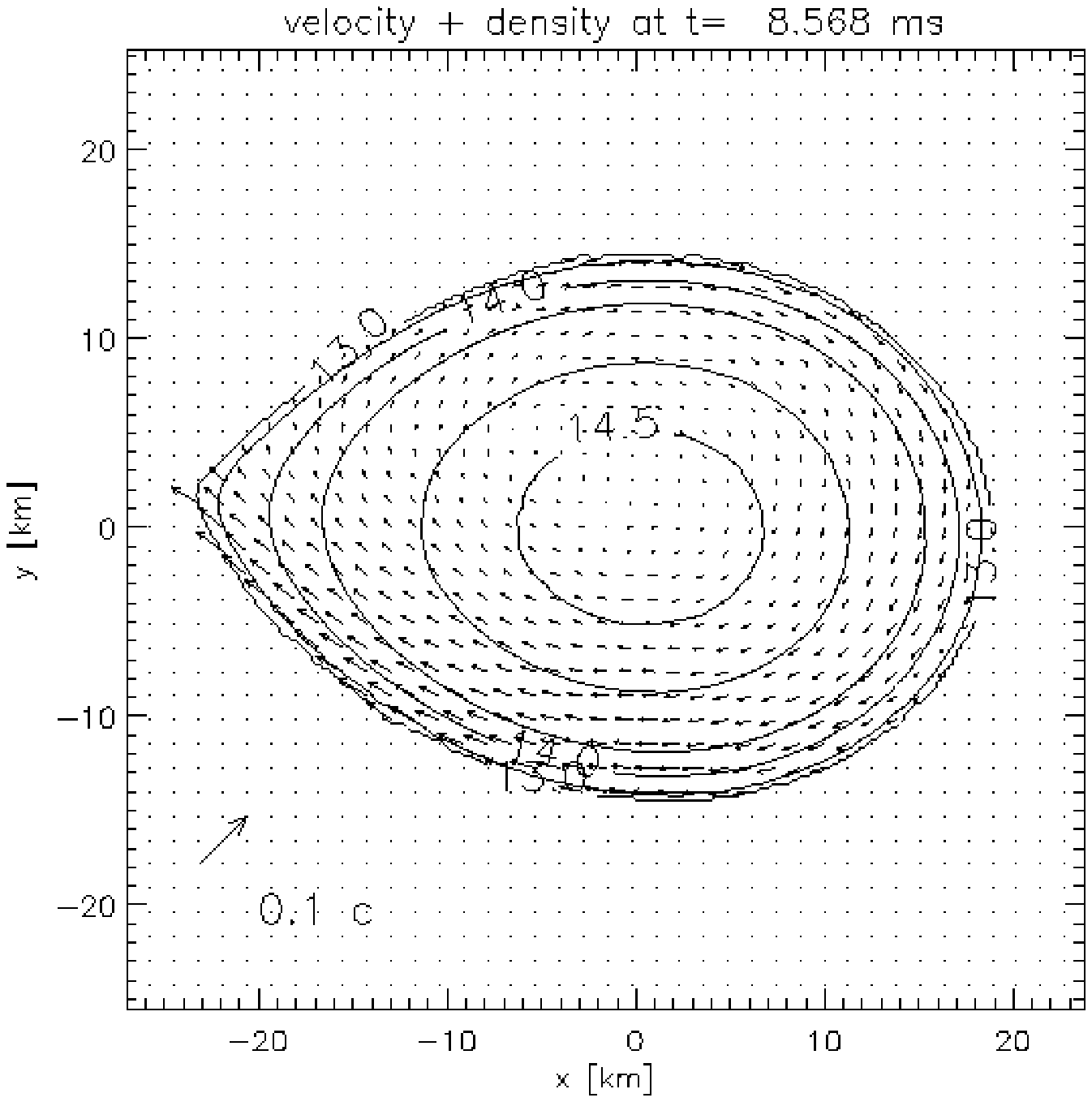,width=11cm,angle=0}
\hspace*{0cm}\psfig{file=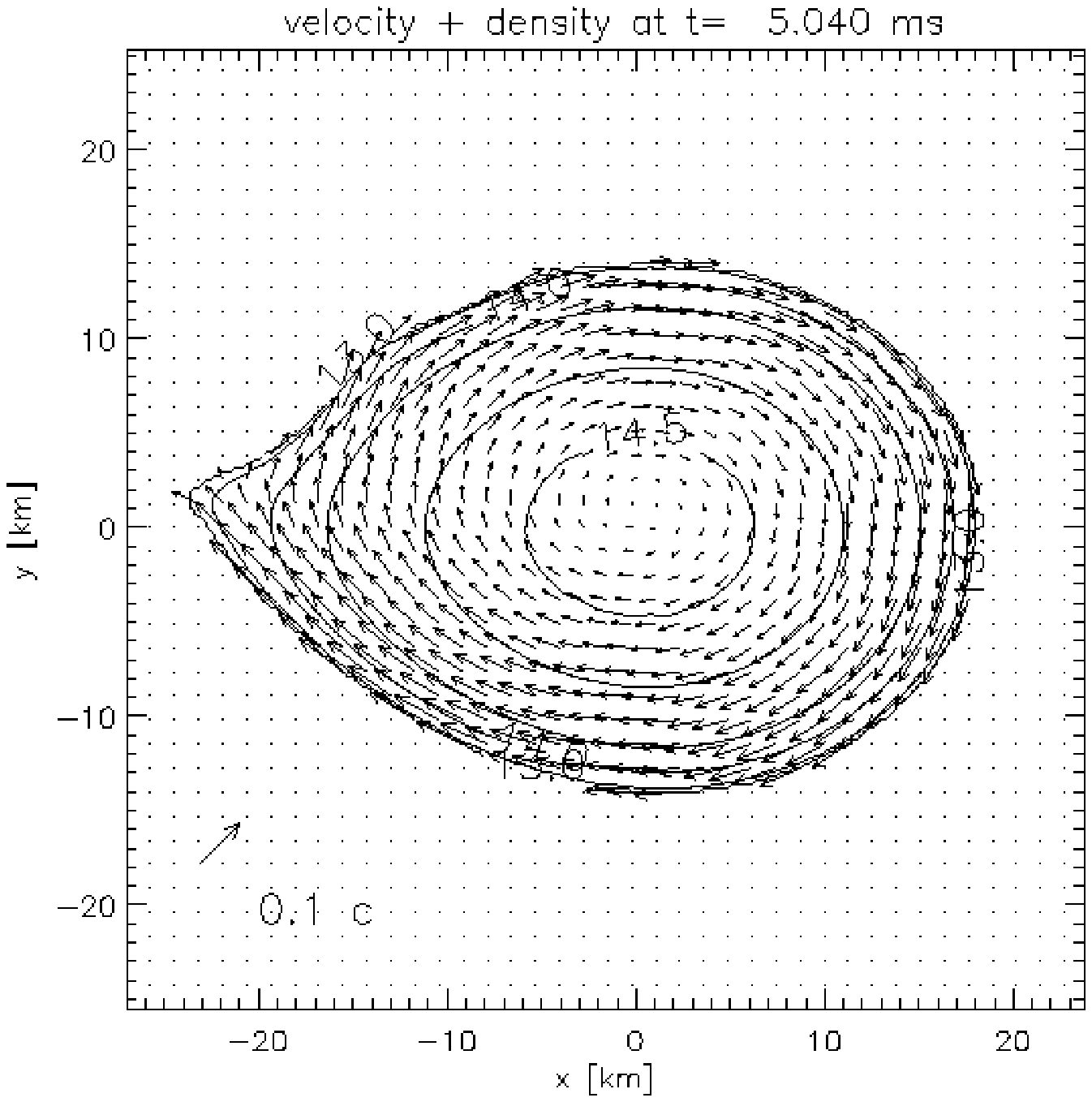,width=11cm,angle=0}

    \caption{\label{vel_onset}Comparison of the velocity fields at the
                              beginning mass transfer.
                              The velocities are plotted in a frame that is 
                              on a circular orbit with a velocity equal to the 
	                      tangential velocity of the star (x-axis through
                              bh and ns, ns at origin). The upper panel
                              refers to a corotating star (run B, q=0.3), the
                              lower one to the case without initial spin (run
                              E, q=0.3).}
\end{figure}

\begin{figure}    
\hspace*{0cm}\psfig{file=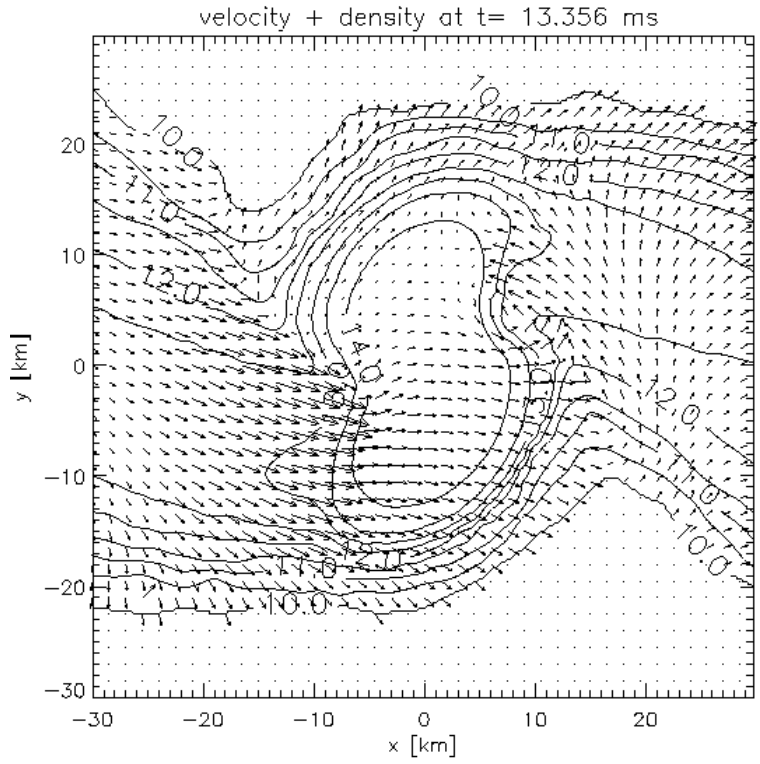,width=11cm,angle=0}
\caption{\label{fission} Fluid motions inside the neutron star induced by
    the mass transfer (same coordiate system as previous Figure). Snapshot
    from run D (no spin, q= 0.1), the contour lines refer to $\log(\rho)$.} 
\end{figure}

% - M_ns(t)
% - rho_max(t)
% - spin_ns(t)/flow patterns
The mass flow into the hole induces a sometimes violent fluid
motion within the neutron star. To illustrate the difference for both spin 
cases, we plot the velocities in a frame that is momentarily rotating on 
a circular orbit (ns and bh both lying on the x-axis) with the neutron star 
being at the origin. In this frame a perfectly corotating binary should 
only exhibit a radial velocity along the x-axis. We show in
Fig. \ref{vel_onset} two cases (corotating and no spin) where mass transfer
has just set in. Note that the matter transferred towards the hole is
coming from the neutron star backside. While the fluid figure of the
corotating case is rather symmetric with respect to the x-axis, the
irrotational case exhibits substantial perturbations on the front side of the
neutron star.\\
The mass transfer induces high-amplitude (with respect to the neutron star 
radius) surface waves travelling with the sense of the orbital motion. They 
whip through the accretion stream connecting the ns with the hole, 
sometimes quenching the accretion stream. An example of such a phase is shown
in Fig. \ref{fission} (same coordinate system as Fig. \ref{vel_onset}).\\
We find that in almost all cases the
core of the neutron star survives all the mass transfer phases and continues
to orbit the hole as a ``mini-neutron star''. Only in our extreme (test-)case
with a black hole of 1.5 \msun\ ($q=0.93$) the neutron star is completely
disrupted to form a hot and massive accretion disk.\\

\subsection{Gravitational Waves}
\begin{figure}    
%\hspace*{0cm}\psfig{file=GW_emission_pmNq01cor_v2.ps,width=7cm,angle=-90}
\hspace*{0cm}\psfig{file=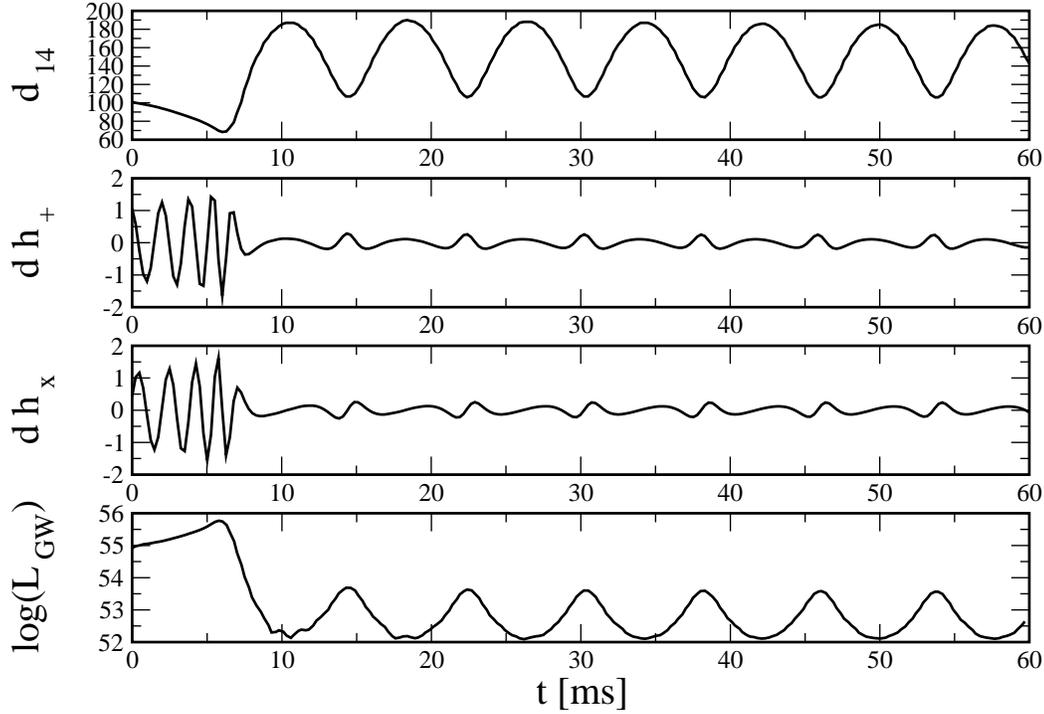,width=12cm,angle=-90}
\caption{\label{GW_pmNq01cor}Gravitational wave emission of run A (q= 0.1,
  corotating neutron star). The upper panel shows the separation of the
  neutron star (defined via a density threshold of $10^{14}$ \gcc) and the
  black hole. Panels two and three show the gravitational amplitudes $h_+$ and
  $h_x$ (d is the distance to the observer) and the last panel gives the
  gravitational wave luminosity. In this run the mini neutron survives the
  whole simulated time, the periastron approaches are clearly visible in the
  gravitational wave luminosity.}
\end{figure}

\begin{figure}    
\hspace*{0cm}\psfig{file=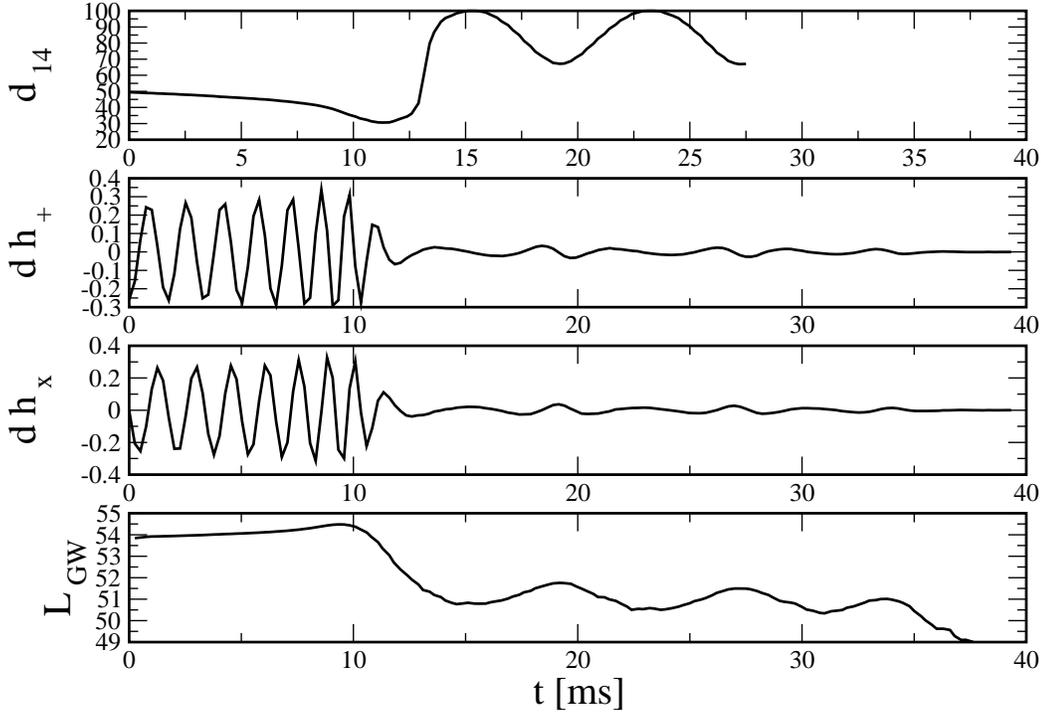,width=12cm,angle=-90}
\caption{\label{GW_pmNq093cor}Same as Fig. \ref{GW_pmNq01cor} for run G. This
  is the only case where the neutron star is completely disrupted. At 27 ms
  the peak density drops below the threshold of $10^{14}$ \gcc (end of curve
  in panel one), at 35 ms the neutron star has completely disappeared and the 
  gravitational wave luminosity drops to essentially zero.}
\end{figure}

\begin{figure}    
\hspace*{0cm}\psfig{file=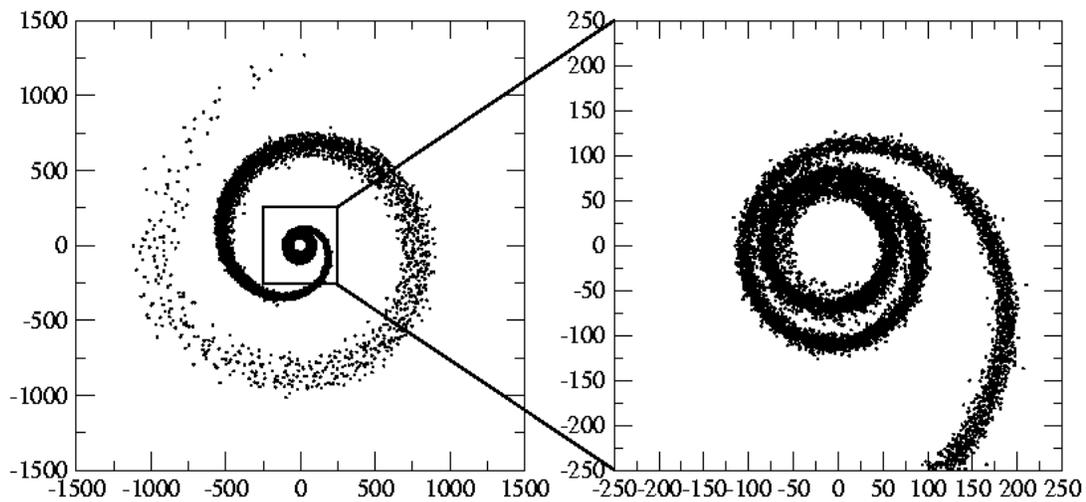,width=12cm,angle=-90}
\caption{\label{Pos_p2}SPH-particle position (at t=18.4 ms) of the
  low-resolution testrun with a soft
  ($\Gamma=2$) polytropic EOS: the neutron star is completely disrupted.}
\end{figure}

\begin{figure}    
\hspace*{0cm}\psfig{file=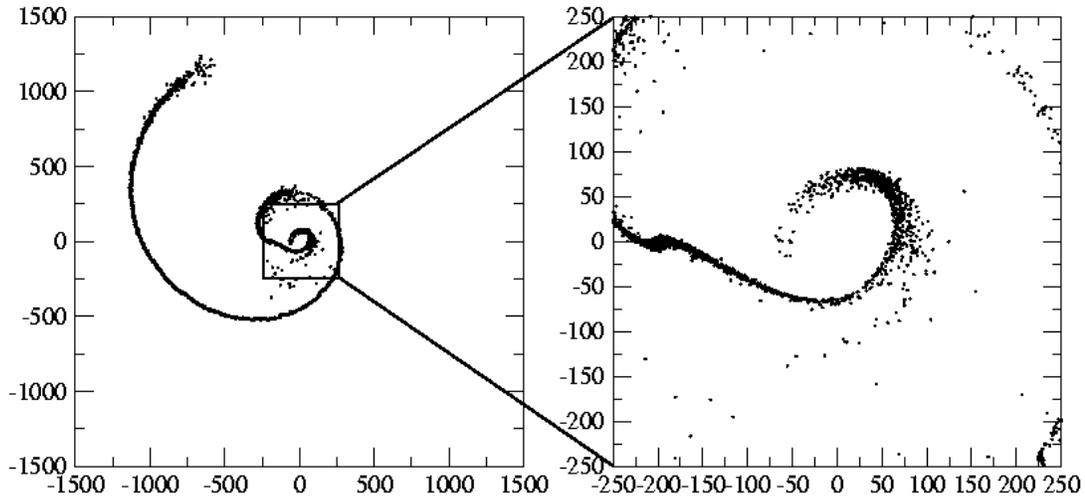,width=12cm,angle=-90}
\caption{\label{Pos_p3}SPH-particle position (at t=18.4 ms) of the
  low-resolution testrun with a stiff
  ($\Gamma=3$) polytropic EOS: a mini-neutron star survives.}
\end{figure}

% Gravitational Waves as Observational Signature of the System Dynamics
% - R_BH_ns(t)
If the more optimistic estimates for the neutron star black hole merger rates
are correct, these systems may be among the first sources to be detected by
ground-based gravitational wave detectors. As will be shown below, the
gravitational wave signal bears a clear imprint of the high-density nuclear
matter equation of state.\\
We show in Figs. \ref{GW_pmNq01cor} and \ref{GW_pmNq093cor} 
the gravitational waves signatures of a run where a mini neutron star survives
(run A) and of a case where the neutron star is finally completely disrupted.
The first panel in each of the figures
shows the mutual separation between the neutron star (defined as the center of
mass of the material with densities in excess of $10^{14}$ \gcc) and the black
hole. After a smooth inspiral phase the mass transfer (directly) into the hole
sets in which increases the orbital separation. The mini neutron star is then
on a non-circular orbit and is further driven back to the hole by the
gravitational wave back-reaction. This sequence of mass transfer, moving out,
coming back in, transferring again mass to the hole etc. continues until the
end of the very long simulation. The existence of the mini neutron star is 
reflected in the gravitational wave luminosity: each periastron approach 
corresponds to a peak in the GW luminosity.\\
In our extreme case, run G, only three more close approaches (apart from the first
one) are visible in the GW-signal. The last encounter tidally disrupts the
neutron star.\\
The survival of the neutron star is a direct consequence of the stiffness
of the EOS. To prove this we have performed two low resolution test runs
with a soft ($\Gamma=2$; run H) and a stiff  ($\Gamma=3$; run I) polytropic
EOS. In the case of the soft EOS the neutron star is completely disrupted (see
Fig. \ref{Pos_p2}) while in the case with the stiff EOS a mini-neutron star
survives (see Fig. \ref{Pos_p3}). Thus, the gravitational wave signal serves
as a direct probe of the poorly known regime of supra-nuclear densities.

\section{Summary and Discussion}
% summary what we have done
%
% Discussion
%           - not as easy as thought/ complex dynamics
%                * GRB: ``not as easy BH+disk''
%                       -> complex interplay 
%		          MT -> widen orbit
%                          GW_BR-> reduce orbital separation
%			  mass-radius-rel: how does star react on mass loss?
%
%		       -> sensitive to q !!!!
%
%                * Disk formation depends on donor spin
%
%                * existence of mini-ns depends on stiffness of EOS
%                  -> visible in GW
%                  -> GW as probe of high-density QCD
We have performed simulations of the accretion dynamics in
neutron star black hole systems. Our simulations used a 3D smoothed particle
hydrodynamics code together with the relativistic mean field nuclear equation
of state of Shen et al. (1998a,1998b) and a detailed neutrino scheme. Gravity
was approximated with Newtonian forces plus back-reaction from the emission of
gravitational waves.\\
Generally we find a rather complex dynamics determined by the interplay between
mass transfer (trying to increase the orbital separation), gravitational wave
backreaction (acting to reduce the separation) and the reaction of the neutron
star to mass loss. This reaction includes both the change of the neutron star
radius and (in some cases) very violent fluid motions inside the star with
amplitudes comparable to the stellar radius. In some of the cases the neutron
star even seems to be close to fissioning into two fragments. Thus, a certain
fraction of systems may result in a black hole orbited by two low-mass 
neutron stars.
We find that many of the analytic estimates that are useful in other 
accreting systems are of limited use for neutron star black hole systems. 
The overall evolution is very sensitive to the neutron star black hole mass 
ratio.\\
We show generally that -if not superimposed by other effects such as a
different tidal deformation or a different tidal lag angle- irrotational donor
stars form easier accretion disks than corotating ones. 
This result is not restricted to neutron star black hole systems, but holds
generally. The reason is that in the irrotational case mass overflowing the 
$L_1$ Lagrange point has more specific angular momentum and therefore a 
larger circularization radius. In the neutron star black hole case more 
material can avoid being fed directly into the Schwarzschild radius if 
the donor star is not spinning.\\
In the context of GRB progenitors neutron star black hole systems are
usually regarded only as a slight modification of the binary neutron star
case. It is often stated that they will yield the ``standard GRB central 
engine'', a
black hole with a hot and massive accretion disk. Here, however, we find it
very difficult to form such a disk. Although somewhat easier in the irrotational
case, we find for all but extreme cases rather low temperature disks ($\sim 3$
MeV) of moderate densities, which are not promising to either launch a GRB via
neutrino annihilation or via energy extraction from the black hole rotation
with suitable magnetic field configurations. Some of our investigated systems
do not form any resolvable disk at all. At the end of the simulation we have
in these cases a binary system of a black hole and a low mass neutron star 
engulfed in a common envelope of low density material that has been tidally 
removed from the neutron star.\\
We find in all but extreme cases that the core of the initial star survives
all mass transfer episodes as a ``mini neutron star''. Test runs with
polytropic equations of state ($\Gamma=2$ and $\Gamma=3$) show that 
in the case of a stiff equation of state a mini-neutron star forms while
a soft EOS leads to a complete tidal disruption of the neutron star.
This result is consistent with the work of Lee/Lee and Kluzniak. Janka et al. (1999) 
also find in some cases that the core of the neutron star survives the initial
mass transfer episodes but is later on completely disrupted. This is
consistent with their EOS being somewhat softer than the one we use (for a
comparison of both EOSs see Figure 5 of Rosswog and Davies (2002)).\\
%added in response to referee's report
The survival of a mini neutron star and the difficulty to form a massive
accretion disk are not unrelated; the surviving stellar core acts as some 
kind of storage that prevents mass from flowing towards the hole. Both effects
are a result of the stiffness of the neutron star equation of state as can be
seen from an inspection of the test runs with a stiff (neutron star core, low
mass disk) and a soft polytropic equation of state (complete disruption, more
massive disk).  Janka et al. (1999) et al. found values for the neutrino 
emission and annihilation that are rather optimistic for GRBs. In both sets 
of simulations different numerical methods are used, but we regard the 
differences resulting from different techniques as minor. We consider the 
uncertainty in the basic physics at supranuclear densities and the sensitivity
of the system dynamics to this part of the EOS to be the decisive factors for 
the differences in the results. There are not many equations of state for the 
purpose of such simulations (temperature dependent, large range of densities,
no $\beta$-equilibrium assumed) available, the used examples of the 
Lattimer-Swesty EOS (simulations of Janka et al. 1999) and the Shen-EOS 
(our simulations) may therefore indicate the possible range of results as a 
function of the poorly known input physics.\\
%end addition
In most cases huge tidal tails of decompressed neutron star material
form. During the decompression phase the adiabatic exponent drops from values
above 3 to values close to 4/3. This goes along with a recombination of
nucleons into nuclei and therefore the release of several MeV per
nucleon. These effects drive a rapid lateral expansion of the tidal tails that
will therefore end up as large, structured disks around the hole.\\
The gravitational wave luminosity reaches peak values close to $10^{56}$ erg/s
at the first encounter. The survival of the mini neutron star 
leaves a clear imprint on the gravitational wave signature. All the subsequent
periastron approaches yield a luminosity peak roughly two orders of magnitude
below the value of the first close approach. In our extreme case with mass
ratio close to one both amplitudes and luminosity die off abruptly once the
mini neutron star is finally disrupted. Thus a softening of the equation of 
state at supranuclear densities due to, for example, hyperon condensates would
be directly observable in the gravitational wave signature of a neutron star
black hole system.

\vspace*{1cm}
{\bf Acknowledgements}\\
\vspace*{0.1cm}\\
It is a pleasure to thank the Leicester supercomputer team Stuart Poulton,
Chris Rudge and Richard West for their excellent support.
Most of the computations reported here were performed using the UK
Astrophysical Fluids Facility (UKAFF). Part of this work has been performed 
using the University of Leicester Mathematical Modelling Centre's
supercomputer which was purchased through the EPSRC strategic equipment 
initiative.\\
This work was supported by a PPARC Rolling Grant for Theoretical Astrophysics.
SR gratefully acknowledges the support via a PPARC Advanced Fellowship.
RS whishes to thank the UK Astrophysical Fluid Facility
(UKAFF) in Leicester for the kind hospitality during a visit, where
parts of this work were accomplished, and he would like to acknowledge
the funding of this visit by the EU FP5 programme.

\begin{appendix}
\section{Gravitational Waves}

We treat gravitational waves in the quadrupole approximation 
(see e.g. Misner et al. 1973). 
The reduced quadrupole moments can be written in terms of SPH-particle 
properties (Centrella and McMillan 1993)
\be
I_{jk}= \sum_i m_i (x_{ji} x_{ki} - \frac{1}{3} \delta_{jk} r_i^2).
\ee
The second time derivatives, $\ddot{I}_{jk}$, can be expressed in terms 
of particle properties by straight forward differentiation. For the third 
order time derivatives we use the analytical derivatives of a cubic spline 
interpolation of the $\ddot{I}_{jk}$. The retarded gravitational wave 
amplitudes for a distant observer along the binary axis at distance $d$
are given by
\be
d \, h_{+}= \frac{G}{c^{4}}(\ddot{I}_{xx}-\ddot{I}_{yy})
\ee
and
\be
d \, h_{\times}= 2 \frac{G}{c^{4}}\ddot{I}_{xy},
\ee
the total gravitational wave luminosity is approximated as
\be
L_{GW}\cong \frac{G}{5c^5} I^{(3)}_{jk} I^{(3)}_{jk},
\ee
where the superscript denotes the third derivative with respect to time.

\end{appendix}

%%%%%%%%%%%%%%%%%%%%%%%%%%%%%%%%%%%%%%%%%%%%%%%%%%%%%%%%%%%%%%%%%%%%%%%%%
%                           Bibliography                                %
%%%%%%%%%%%%%%%%%%%%%%%%%%%%%%%%%%%%%%%%%%%%%%%%%%%%%%%%%%%%%%%%%%%%%%%%%

%%%%%%%%%%%%%%%%%%%%%%%%%%%%%%%%%%%%%%%%%%%%%%%%%%%%%%%%%%%%%%%%%%%%%%%%%%%%%
%
%                    PUT FIGURES HERE !           
%
%%%%%%%%%%%%%%%%%%%%%%%%%%%%%%%%%%%%%%%%%%%%%%%%%%%%%%%%%%%%%%%%%%%%%%%%%%%%%
\clearpage
\clearpage
\clearpage
\clearpage
%%%%%%%%%%%%%%%%%%%%%%%%%%%%%%%%%%%%%%%%%%%%%%%%%%%%%%%%%%%%%%%%%%%%%%%
%                     pmNq01cor                                       %
%%%%%%%%%%%%%%%%%%%%%%%%%%%%%%%%%%%%%%%%%%%%%%%%%%%%%%%%%%%%%%%%%%%%%%%
%%%%%%%%%%%%%%%%%%%%%%%%%%%%%%%%%%%%%%%%%%%%%%%%%%%%%%%%%%%%%%%%%%%%%%%%
%                     pmNq03irr                                        %
%%%%%%%%%%%%%%%%%%%%%%%%%%%%%%%%%%%%%%%%%%%%%%%%%%%%%%%%%%%%%%%%%%%%%%%%
%%%%%%%%%%%%%%%%%%%%%%%%%%%%%%%%%%%%%%%%%%%%%%%%%%%%%%%%%%%%%%%%%%%%%%%%%%%
%                     pmNq093cor                                          %
%%%%%%%%%%%%%%%%%%%%%%%%%%%%%%%%%%%%%%%%%%%%%%%%%%%%%%%%%%%%%%%%%%%%%%%%%%%
\clearpage
\clearpage
\clearpage

\clearpage
\clearpage
\clearpage
\clearpage
\clearpage

\clearpage

\clearpage

\clearpage
\end{document}